\documentclass{article}

\usepackage{arxiv}

\usepackage[utf8]{inputenc} % allow utf-8 input
\usepackage[T1]{fontenc}    % use 8-bit T1 fonts
\usepackage{hyperref}       % hyperlinks
\usepackage{url}            % simple URL typesetting
\usepackage{booktabs}       % professional-quality tables
\usepackage{amsfonts}       % blackboard math symbols
\usepackage{nicefrac}       % compact symbols for 1/2, etc.
\usepackage{microtype}      % microtypography
\usepackage{lipsum}		% Can be removed after putting your text content
\usepackage{graphicx}
\usepackage[authoryear]{natbib}
\usepackage{doi}

\usepackage{mathtools,upgreek,eucal,mathrsfs,enumitem,amsthm,amsmath}
\usepackage{times,epsfig,makeidx,amsfonts,amsmath,dcolumn,multirow,amssymb,colortbl,bm,url}
\usepackage{algorithm}
\usepackage[]{algorithmic}
 \usepackage[flushleft]{threeparttable}

\usepackage{caption}
\usepackage{subcaption}

\newtheorem{theorem}{Theorem}

\DeclareMathOperator{\arcsinh}{arcsinh}

\newcommand{\diag}{\mathop{\mathrm{diag}}}

\newcommand{\beq}{\begin{equation}}
\newcommand{\eeq}{\end{equation}}

\newcommand{\bbeta}{\bm{\beta}}
\newcommand{\ff}{\textbf{f}}

\newcommand{\defn}{\begin{quote}{\bf Definition. }}
\newcommand{\edefn}{\end{quote}}
\newcommand{\thm}{\begin{theorem}}
\newcommand{\ethm}{\end{theorem}}

\newcommand{\bmat}[1]{\left ( \begin{array}{#1}}
\newcommand{\emat}{\end{array}\right )}

\newcommand{\ts}{^{\sf T}}
\newcommand{\tstar}{^{*\sf T}}
\newcommand{\tst}{'^{\sf T}}

\newcommand{\x}{ \textbf{x} }

\newcommand{\w}{ \textbf{w} }
\newcommand{\V}{ \textbf{V} }

\theoremstyle{definition}

\title{A Unifying Framework for Flexible Excess Hazard Modeling with Applications in Cancer Epidemiology}
\author{ \href{https://orcid.org/0000-0000-0000-0000}{\includegraphics[scale=0.06]{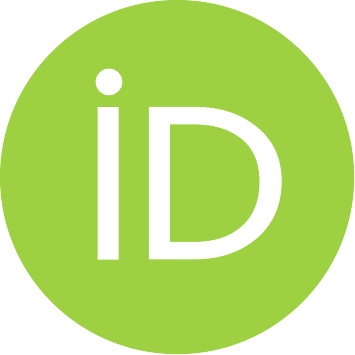}\hspace{1mm}Alessia Eletti}\\
	Department of Statistical Science\\
	University College London \\
	London, UK\\
	\texttt{alessia.eletti.19@ucl.ac.uk}  \\
	\And
	\href{https://orcid.org/0000-0002-9010-2646}{\includegraphics[scale=0.06]{orcid.pdf}\hspace{1mm}Giampiero Marra}\\
	Department of Statistical Science\\
	University College London \\
	London, UK\\
	\texttt{giampiero.marra@ucl.ac.uk}  \\
		\And
	\href{https://orcid.org/0000-0002-5290-279X}{\includegraphics[scale=0.06]{orcid.pdf}\hspace{1mm}Manuela Quaresma}\\
Department of Non-Communicable Diseases Epidemiology\\
London School of Hygiene \& Tropical Medicine\\
London, UK.\\
	\texttt{manuela.quaresma@lshtm.ac.uk} \\
	\And
	\href{https://orcid.org/0000-0002-6316-3961}{\includegraphics[scale=0.06]{orcid.pdf}\hspace{1mm}Rosalba radice}\\
Faculty of Actuarial Science and Insurance\\
Business School\\
City, University of London
London, UK.\\
	\texttt{rosalba.radice@city.ac.uk} \\
	\And
	\href{https://orcid.org/0000-0001-7183-8407}{\includegraphics[scale=0.06]{orcid.pdf}\hspace{1mm}Francisco Javier Rubio} \\
	Department of Statistical Science\\
	University College London \\
	London, UK\\
	\texttt{f.j.rubio@ucl.ac.uk} 
	}

\hypersetup{
pdftitle={XXX},
pdfsubject={stat.ME, stat.AP},
pdfauthor={Eletti et al.},
}

\begin{document}
\maketitle

\begin{abstract}
Excess hazard modeling is one of the main tools in population-based cancer survival research. Indeed, this setting allows for direct modeling of the survival due to cancer even in the absence of reliable information on the cause of death, which is common in population-based cancer epidemiology studies. We propose a unifying link-based additive modeling framework for the excess hazard that allows for the inclusion of many types of covariate effects, including spatial and time-dependent effects, using any type of smoother, such as thin plate, cubic splines, tensor products and Markov random fields. In addition, this framework accounts for all types of censoring as well as left-truncation. Estimation is conducted by using an efficient and stable penalized likelihood-based algorithm whose empirical performance is evaluated through extensive simulation studies. Some theoretical and asymptotic results are discussed. Two case studies are presented using population-based cancer data from patients diagnosed with breast (female), colon and lung cancers in England. The results support the presence of non-linear and time-dependent effects as well as spatial variation. The proposed approach is available in the \texttt{R} package \texttt{GJRM}.
\end{abstract}

% keywords can be removed

%----------------------------------------------------------------------------------------------------------------------------------------------------------------------

\section{Introduction}
\label{intro}

One of the aims of population cancer epidemiology consists of quantifying the survival due to cancer and to describe inequalities in cancer survival outcomes. 
This includes comparisons of cancer survival between different subgroups of the populations, such as those defined by different socio-economic or geographic factors.
Cancer survival is typically used as a proxy for the overall effectiveness of the healthcare system in the treatment and management of cancer \citep{coleman:2014}, and it is increasingly used to formulate cancer control strategies \citep{DofH:2011}. Data for cancer research are available from population-based cancer registries which collect a standard set of information for every cancer registration, covering patient demographics, tumor characteristics and type of treatment. Many efforts have been made in recent years to augment cancer registration data with relevant clinical information contained in other electronic health databases. Such enriched data create new opportunities for more complex cancer research questions to be investigated.

There are three main frameworks for analyzing survival data. The first is the overall survival framework, where all-cause mortality is studied. This quantity is not of interest in cancer survival studies because it does not quantify the survival due to cancer. The second is the cause-specific framework, where information on the different causes of death is available, e.g. in the death certificates. This addresses the previous issue as it indeed accounts for the different causes of mortality in the population. Unfortunately, death certificates are unreliable in virtually any country in the world, at least at the population level. The third is the relative survival framework, which can be formulated in the absence of information on the cause of death. In this framework, the idea is to separate the hazard associated to other causes of death from that associated to cancer.
This is done by assuming an additive decomposition of the individual hazard function, $h(\cdot)$, into two parts: the hazard associated to other causes of death, $h_O(\cdot)$, and the hazard associated to cancer, $h_E(\cdot)$ (\citealp{esteve:1990}):
\begin{eqnarray}\label{eq:excess_hazard}
h(t \mid\x) = h_O(\text{age} + t) + h_E(t\mid\x),
\end{eqnarray}
where ``$\text{age}$'' is the age at diagnosis of cancer and \x represents the available patient characteristics. The hazard associated to other causes of death, $h_O(\text{age} + t)$, is typically replaced by the population hazard rate $h_P(\text{age}+t\mid \w)$, which is obtained from life tables based on available characteristics denoted by the generic vector $\w \subset \x$ which can possibly include, in addition to age at death or censoring ($\text{age}+t$), gender and calendar year, socio-economic status, ethnicity or region of residence \citep{rachet:2015}. More specifically, $h_O( \text{age} + t)$ represents the true theoretical hazard function associated to other causes of death and as such it is unknown in practice. For this reason it is approximated by $h_P( \text{age} + t \mid \w)$, which can instead be extracted from national life tables as mentioned above. As an aside, note that the true theoretical hazard function $h_O( \text{age} + t)$ may depend on \textbf{x}, or on a subset of \textbf{x}, or even on covariates that are not recorded.

The hazard associated to cancer, $h_E(t\mid\x)$, is often referred to as the \textit{excess hazard}. The excess hazard function is typically modeled using the available patient characteristics, denoted by $\x$ which can, for instance, incorporate continuous and categorical variables in our framework. Several approaches for estimating the excess hazard have been explored in the literature, such as non-parametric methods, which aim at estimating the cumulative excess hazard \citep{perme:2012} and the net survival \citep{perme2009approach, perme:2016, pavlivc:2019}, parametric methods based on flexibly modeling the baseline excess hazard or cumulative hazard using splines \citep{charvat:2016,Cramb:2016,fauvernier:2019,lambert:2009,quaresma:2019}, and modeling the baseline excess hazard function using flexible parametric distributions \citep{rubio:2019S}. Most approaches assume a proportional hazards (PH) structure (with the option of adding time-dependent effects as originally proposed by \cite{cox:1972}, which is a convenient way of bypassing the proportionality assumed by the PH setting), with the exception of \cite{rubio:2019S}, who adopt a general hazard structure that contains the PH, accelerated hazards, and the accelerated failure time (AFT) models as particular cases. 

We propose a flexible parametric modeling framework. In this respect, it should be noted that non-parametric and parametric approaches are generally viewed as complementary by practitioners, rather than mutually exclusive. This view is strengthened by the fact that they are not directly comparable. The interpretation for non-parametric models is, in fact, different than for parametric models. This includes but is not limited to the fact that parametric approaches can account for covariates directly while non-parametric approaches cannot \citep{perme:2012}. Further, as mentioned above the available approaches do not allow one to model the excess hazard function, as they represent estimates of the cumulative hazard or net survival \citep{perme:2012}. Instead, parametric approaches allow one to estimate and plot the excess hazard function as this function is explicitly available \citep{rubio:2019S}.

Finally, with regard to our choice of taking a parametric approach, we note that Cox has encouraged the broader use of parametric survival models for empirical modeling \citep{Reid,Hjort}. This is because they facilitate model estimation and comparison, easily allow for the calculation and visualization of, for instance, the estimated baseline hazard and survival functions, and allow one to calculate many quantities of interest and their related intervals (e.g., time-dependent hazard or odds ratios). Moreover, we overcome the generally restrictive nature of traditional parametric models by proposing a splines-based framework which allows for a great degree of modeling flexibility.

Based on the decomposition of the hazard function \eqref{eq:excess_hazard}, the cumulative hazard function can be written as
\begin{equation}\label{eq:excess_chazard}
\begin{aligned}
H(t\mid\x) &= \int_0^{t} h(r\mid \x)dr = H_P(\text{age} + t\mid\w) - H_P(\text{age}\mid\w) + H_E(t\mid\x).
\end{aligned}
\end{equation}
Consequently, the survival function can be factorized as follows:
\begin{equation}\label{eq:net_surv}
\begin{aligned}
S(t\mid\x) &= \exp\left\{-H(t\mid\x)\right\} = \exp\left\{ - H_P(\text{age} + t\mid\w) + H_P(\text{age}\mid\w)\right\}\exp\left\{-H_E(t\mid\x)\right\}.
\end{aligned}
\end{equation}
The survival function associated to the excess hazard, $S_N(t\mid\x) = \exp\left\{-H_E(t\mid\x)\right\}$, is denoted as the (individual) \textit{net survival}. The concept of net survival is usually favored by international agencies and programs devoted to the study of cancer epidemiology, as well as policy-makers, as it is not affected by other causes of mortality, under the assumed model \eqref{eq:excess_hazard}; we refer the reader to \cite{rubio:2019S} and \cite{rubio:2019B} for a discussion on these points.

Building on \cite{Marra2019}, we present a flexible methodology that is capable of handling simultaneously all types of censoring as well as left-truncation, while accounting for the excess hazard. Often only right-censoring and potentially left-truncation is allowed \citep[e.g.,][]{fauvernier:2019, quaresma:2019}, thus accounting for any type of censoring broadens the applicability of our framework. Further, a variety of covariate effects, including time-dependent effects, can be flexibly estimated via additive predictors with several types of smoothers. Our framework can also accommodate spatial effects in the definition of the additive predictor of an excess hazard model, a feature that is not available in other frameworks and software; this allows us to explore geographic disparities in cancer survival. The proposed model yields as special cases the widely used PH model, which allows for the usual Cox-like interpretation of the estimated effects, as well as the proportional odds (PO) model. The framework is based on modeling transformations of the survival function which we found to perform well in practice. An advantage of using this scale is that the post-estimation extraction of the (sub-)population net survival, a quantity often of interest to practitioners, is notably quicker when compared to approaches which model on the hazards scale; the need for numerical integration in the latter implies a higher computational time.

The resulting additive model is very flexible since the baseline hazard is modeled by means of monotonic P-splines. This is more efficient and parsimonious than using a non-parametric hazard, as in the Cox model, and it is more flexible than strong parametric assumptions such as those in AFT models. Parameter estimation is based on a penalized maximum likelihood approach that allows for stable and efficient computations where smoothness is guaranteed by means of a quadratic penalty. In order to allow for transparent and reproducible research as well as faster dissemination of scientific results in industry and academia, the proposed modeling framework is implemented in the \texttt{GJRM R} package \citep{GJRM}. This implementation allows the applied end-user to obtain and visualize relevant quantities such as population net survival and excess hazard, and their confidence intervals, and easily perform model comparisons. Various examples of code usage can be found in the on-line Supplementary Material as well as on the public repository {\small\texttt{https://github.com/FJRubio67/LBANS/}}, where two publicly available datasets are analyzed.

Sections \ref{model.gen1} and \ref{pll} present the model formulation and the model's penalized log-likelihood. Section \ref{estimation-inference} discusses parameter estimation and inference as well as some theoretical results. Section \ref{simulation} contains the results of the simulation study. Sections \ref{appl1} and \ref{appl2} present two case studies in the context of cancer epidemiology. Section \ref{discussion} concludes the paper with a discussion and potential directions for future research. Finally, for the sake of space, several details are collected in the on-line Supplementary Material.

\section{Flexible excess hazard model} \label{model.gen1}

For individual $i = 1, \ldots, n$, where $n$ represents the sample size, let $T_i$ denote the true event time and have a conditional net survival function denoted by $S_N(t_{i} \mid \textbf{x}_{i}; \bbeta)= \exp\left\{ - H_E(t_{i} \mid \textbf{x}_{i}; \bbeta) \right\}\in (0,1)$, where $\textbf{x}_i$ represents a generic vector of patient characteristics that has an associated regression coefficient vector $\bbeta \in \mathbb{R}^w$, where $w$ is the length of $\bbeta$. A link-based additive net survival model can be written as
\beq
g\left\{S_N(t_{i}\mid\textbf{x}_{i}; \bbeta)\right\} = \eta_{i}(t_{i},\textbf{x}_{i};\ff(\bbeta)),
\label{g.fun}
\eeq
where $g:(0,1) \rightarrow \mathbb{R}$ is a monotone and twice continuously differentiable link function with bounded derivatives and hence invertible, $\eta_{i}(t_{i},\textbf{x}_{i};\ff(\bbeta)) \in \mathbb{R}$ is an additive predictor which includes a baseline function of time, or a stratified set of functions of time, and several types of covariate effects (see the next section), and $\ff(\bbeta)$ is a vector function of $\bbeta$ whose main role is to impose the monotonicity constraint, discussed in Section \ref{pll}, needed when evaluating the baseline function of time contained in the additive predictor. Note that the choice for $g$ determines the scale of the analysis \citep[e.g.,][]{Liu}. %Also, an important benefit of modeling on the survival scale is that it eliminates the need for numerical integration. 

Rearranging (\ref{g.fun}) yields $S_N(t_{i}\mid\textbf{x}_{i}; \bbeta) =G\left\{\eta_{i}(t_{i},\textbf{x}_{i};\ff(\bbeta))\right\}$, where $G$ is an inverse link function. The cumulative hazard and hazard functions, $H$ and $h$, are defined as $H_E(t_{i}\mid\textbf{x}_{i}; \bbeta) = -\log\left[G\left\{\eta_{i}(t_{i},\textbf{x}_{i};\ff(\bbeta))\right\}\right]$ and
\beq
h_E(t_{i}\mid\textbf{x}_{i}; \bbeta) = -\frac{G'\left\{\eta_{i}(t_{i},\textbf{x}_{i};\ff(\bbeta))\right\}}{G\left\{\eta_{i}(t_{i},\textbf{x}_{i};\ff(\bbeta))\right\}}\frac{\partial \eta_{i}(t_{i},\textbf{x}_{i};\ff(\bbeta))}{\partial t_{i}},
\label{h.fun}
\eeq 
where $G'\left\{\eta_{i}(t_{i},\textbf{x}_{i};\ff(\bbeta))\right\}=\partial G\left\{\eta_{i}(t_{i},\textbf{x}_{i};\ff(\bbeta))\right\}/\partial \eta_{i}(t_{i},\textbf{x}_{i};\ff(\bbeta))$. Table \ref{marl} displays the functions $g$, $G$ and $G'$ considered in this work.

\begin{table}[htbp]
\footnotesize
\begin{center}
\begin{tabular}{lcccc}
\hline
Model & Link $g(S)$ & Inverse link $g^{-1}(\eta)=G(\eta)$ & $G'(\eta)$ \\ \hline
Prop. hazards (\texttt{''PH''}) & $\log \left\{-\log(S) \right\}$  & $\exp\left\{-\exp(\eta)\right\}$ &  $-G(\eta)\exp(\eta)$ \\
Prop. odds (\texttt{''PO''}) & $-\log\left(\frac{S}{1-S}\right)$ & $\frac{\exp(-\eta)}{1+\exp(-\eta)}$ & $-G^2(\eta)\exp(-\eta)$  \\
probit (\texttt{''probit''}) & $-\Phi^{-1}(S)$ & $\Phi(-\eta)$ & $-\phi(-\eta)$\\
\hline
\end{tabular}
\end{center}
\caption{Functions implemented in \texttt{GJRM} (see \cite{Marra2019} and references therein). $\Phi$ and $\phi$ are the cumulative distribution and density functions of a univariate standard normal distribution. The first two functions are typically known as log-log and -logit links, respectively.}
\label{marl}
\end{table}

\subsection{Additive predictor}\label{ap}

For the sake of simplicity, the dependence on covariates and parameters has been dropped when discussing the construction of $\eta_{i}$. Since $t_i$ can be treated as a regressor, we define an overall covariate vector $\textbf{z}_{i}$ made up of $\textbf{x}_{i}$ and $t_{i}$. An additive predictor allows for various types of covariate effects as well as their flexible functional form determination. An additive predictor is defined as
\beq
\eta_{i}=\beta_{0}+\sum_{k=1}^{K} s_{k}(\textbf{z}_{k i}), \quad i=1,\ldots,n,
\label{linpred}
\eeq
where $\beta_{0}\in\mathbb{R}$ is an overall intercept, $\textbf{z}_{k i}$ denotes the $k^{th}$ sub-vector of the complete vector $\textbf{z}_{i}$ and the $K$ functions $s_{k}(\textbf{z}_{k i})$ denote effects which are chosen according to the type of covariate(s) considered. Each $s_{k}(\textbf{z}_{k i})$ can be represented as a linear combination of $J_{k}$ basis functions $b_{k j_{k}}(\textbf{z}_{k i})$ and regression coefficients $f_{k j_k}(\beta_{k j_{k}}) \in \mathbb{R}$, that is \citep[e.g.,][]{Wood}
\beq
\sum_{j_{k}=1}^{J_{k}} f_{k j_k}(\beta_{k j_{k}}) b_{k j_{k}}(\textbf{z}_{k i}).
\label{bfunc}
\eeq The above formulation implies that the vector of evaluations $\left\{s_{k}(\textbf{z}_{k 1}), \ldots, s_{k}(\textbf{z}_{k n})\right\}\ts$ can be written as $\textbf{Z}_k \ff_k(\bbeta_k)$ with $\ff_k(\bbeta_k) = (f_{k1}(\beta_{k1}), \ldots , f_{k J_k}(\beta_{k J_k}) )\ts$ and design matrix $\textbf{Z}_k[i, j_k] = b_{k j_k}(\textbf{z}_{k i})$. This allows the predictor in equation (\ref{linpred}) to be written as
\beq
\bm\eta=\beta_{0} \textbf{1}_n + \textbf{Z}_{1} \ff_1(\bm \beta_{1}) + \ldots + \textbf{Z}_{K} \ff_K( \bm \beta_{K}),
\label{linpredv}
\eeq
where $\textbf{1}_n$ is an $n$-dimensional vector made up of ones. Equation (\ref{linpredv}) can also be written in a more compact way as $\bm\eta=\textbf{Z} \ff(\bm \beta)$, where $\textbf{Z}=(\mathbf{1}_n,\textbf{Z}_{1},\ldots,\textbf{Z}_{K})$ and $\ff(\bm \beta) = (\beta_{0}, \ff(\bm \beta_{1})\ts, \ldots, \ff(\bm \beta_{K})\ts)\ts$. Additional observations on the additive predictor described here can be found in on-line Supplementary Material .

Each $\bm \beta_{k}$ has an associated quadratic penalty $\lambda_{k} \bm \beta_{k}\ts \textbf{D}_{k} \bm \beta_{k}$, used in fitting, whose role is to enforce specific properties on the $k^{th}$ function, such as smoothness. Note that matrix $\textbf{D}_{k}$ only depends on the choice of the basis functions. The smoothing parameter $\lambda_{k} \in [0,\infty)$ controls the trade-off between fit and smoothness, and hence determines the shape of the estimated smooth function. The overall penalty can be defined as $\bm \beta\ts \textbf{S}\bm \beta$, where $\textbf{S} =\diag(0,\lambda_{1}\textbf{D}_{1}, \ldots, \lambda_{K}\textbf{D}_{K})$. Note that smooth functions are subject to centering (identifiability) constraints which can be imposed as described in \cite{Wood}. Depending on the types of covariate effects one wishes to model, several definitions of basis functions and penalty terms are possible. Examples include thin plate, cubic and P- regression splines, tensor products, Markov random fields (MRFs), random effects, Gaussian process smooths \citep[see][for all the options available]{Wood}. More details can be found in the case studies reported in Section \ref{sec:applications}.

Finally, observe that in (\ref{h.fun}) quantity $\partial \eta_{i}(t_{i},\textbf{x}_{i};\ff(\bbeta))/\partial t_{i}$ is required. Re-writing $\eta_{i}(t_{i},\textbf{x}_{i};\ff(\bbeta))$ as $\textbf{Z}_{i}(t_{i},\textbf{x}_{i})\ts \ff(\bbeta)$, the derivative of interest can be obtained as $\underset{\varepsilon\rightarrow 0}{\lim}  \left\{\frac{\textbf{Z}_{i}(t_{i}+\varepsilon,\textbf{x}_{i})-\textbf{Z}_{i}(t_{i}-\varepsilon,\textbf{x}_{i})}{2\varepsilon}\right\}\ts \ff(\bbeta) = \textbf{Z}_{i}\tst\ff(\bbeta)$, where, depending on the type of spline basis employed, $\textbf{Z}'_{i}$ can be calculated either by a finite-difference method or analytically.

\section{Penalized log-likelihood} \label{pll}

The unifying framework proposed in this paper supports excess hazard modeling, all types of censoring and left-truncation in addition to the flexible additive predictor introduced in Section \ref{ap}. As the case studies presented in Section \ref{sec:applications} involve excess hazard modeling on right-censored data, which is the most common scenario in cancer research, here we will define the setting and the log-likelihood only for this case. A detailed discussion of the full log-likelihood for the general case can be found in on-line Supplementary Material, while its derivation is reported in on-line Supplementary Material .

When the $i^{th}$ true event time $T_i$ is known exactly, the individual is said to be uncensored. In some cases, however, $T_i$ may only be known to be larger than a certain time $R_i$, in which case the individual is said to be right-censored and $R_i$ is the random right-censoring time. The censoring type of the $i^{th}$ observation can be summarized through the use of the indicator functions $\delta_{Ri}$ and $\delta_{Ui}$, where $\delta_{Ri} = 1$ if the observation is right-censored and 0 otherwise while $\delta_{Ui} = 1$ if it is uncensored and 0 otherwise.

Let us assume that a random \textit{i.i.d.} sample $\left\{(r_{i}, \delta_{Ui}, \delta_{Ri}, \textbf{x}_{i})\right\}_{i=1}^n$ is available, where $r_i$ is either the time of death or the observed right-censoring time, and that censoring is independent and non-informative conditional on $\textbf{x}_{i}$. Let us also write $S_N(t_i \mid \x_i) = S_N\left\{\eta_{i}(t_{i})\right\}$ in order to make the dependence of the net survival on $\eta$ explicit. The log-likelihood function associated to the additive excess hazard model \eqref{eq:excess_hazard}--\eqref{eq:net_surv} can be written as 
\begin{equation}\label{likequat}
\begin{aligned}
\ell(\bbeta) = \sum_{i=1}^n  &\delta_{Ui}\log\left[ h_P(\text{age}_i + r_i\mid \w_i)  S_N\left\{\eta_{i}(r_{i})\right\}  - \frac{\partial S_N\left\{\eta_{i}(r_{i})\right\}}{\partial \eta_{i}(r_{i})} \frac{\partial \eta_{i}(r_{i})}{\partial r_{i}}  \right] \\
+ \sum_{i=1}^n & \delta_{Ri}\log\left[S_N\left\{\eta_{i}(r_i)\right\} \right] + C_i ,
\end{aligned}
\end{equation}
where $r_i$ is the exact event time when $\delta_{U i} = 1$ and where $C_i$ is a constant with respect to the model's parameters whose expression can be found in on-line Supplementary Material . 

The proposed model allows for a high degree of flexibility, which is why penalized estimation of $\bbeta$ is advisable. In order to prevent over-fitting, we maximize the penalized log-likelihood
\begin{equation}\label{Penloglik}
    \ell_p(\bbeta)= \ell(\bbeta) -\frac{1}{2} \bbeta\ts \textbf{S} \bbeta.
\end{equation}
To ensure that the estimated survival function is monotonically decreasing or equivalently that the hazard function is positive, the time effects are modeled using the monotonic P-spline approach. Let $s(t_{i})= \sum_{j=1}^{J} f_{j}(\beta_j) b_{j}(t_{i})$, where the $b_{j}$ are B-spline basis functions of at least second order built over the interval $[a,b]$, based on equally spaced knots, and the $f_{j}(\beta_j)$ are spline coefficients. A sufficient condition for $s'(t_{i}) \geq 0$ over $[a, b]$ is that $f_{j}(\beta_j) \geq f_{j-1}(\beta_{j-1}), \forall j$ \citep[e.g.,][]{LEI}. Such condition can be imposed by defining the vector function of $\bbeta$ as $\ff(\bbeta) = \bm\Sigma \left\{\beta_{1},\exp(\beta_{2}),\ldots,\exp(\beta_{J})\right\}\ts  $, where $\bm\Sigma[\iota_{1},\iota_{2}] = 0$ if $\iota_{1}<\iota_{2}$ and $\bm\Sigma[\iota_{1},\iota_{2}] = 1$ if $\iota_{1}\geq \iota_{2}$, with $\iota_{1}$ and $\iota_{2}$ denoting the row and column entries of $\bm\Sigma$, and $\bbeta\ts=(\beta_{1},\beta_{2},\ldots,\beta_{J})$ is the parameter vector to estimate. Note that in practice $\bm \Sigma$ is absorbed into the design matrix containing the B-spline basis functions \textbf{Z}. When setting up the penalty term we penalize the squared differences between adjacent $\beta_{j}$, starting from $\beta_{2}$, using $\textbf{D}=\textbf{D}\tstar\textbf{D}^*$ where $\textbf{D}^*$ is a $(J - 2) \times J$ matrix made up of zeros except that $\textbf{D}^*[\iota,\iota + 1] = -\textbf{D}^*[\iota,\iota + 2] = 1$ for $\iota = 1, . . . , J - 2$ \citep{Pya}.

\section{Parameter estimation and inference} \label{estimation-inference}

The estimation approach employed in this article is based on analytical derivative information which helps enhance numerical stability and speed. It is worth noting that, given the structure of (\ref{likequat}), deriving such quantities has been a tedious task. Further, the above mentioned re-parametrization implies a non-linear dependence of $\ff(\bbeta)$ from $\boldsymbol{\beta}$ which additionally complicates the structure of the derivatives, in particular those of the additive predictor $\eta_{i}(t_{i}, \textbf{x}_i; \ff(\bbeta))$ with respect to parameter vector $\boldsymbol{\beta}$. These appear repeatedly in the score and in the Hessian and are given by $\partial \eta_{i}(t_{i}, \textbf{x}_i; \ff(\bbeta))/\partial \boldsymbol{\beta} = \textbf{Z}_i \circ \textbf{E}$, $\partial^2 \eta_{i}(t_{i}, \textbf{x}_i; \ff(\bbeta))/\partial \boldsymbol{\beta} \partial t_{i} = \textbf{Z}_i' \circ \textbf{E}$ and $\partial^2 \eta_{i}(t_{i}, \textbf{x}_i; \ff(\bbeta))/\partial \boldsymbol{\beta}^2 = \diag(\textbf{Z}_i) \circ \Bar{\Bar{\textbf{E}}}$, where $\textbf{Z}_i$ is the transformed covariate vector corresponding to the $i^{th}$ observation, $\circ$ is the Hadamard product, $\diag(\textbf{v})$ is a diagonal matrix with \textbf{v} its diagonal, \textbf{E} is a $\sum_{k=1}^K J_k \times 1$ vector such that its $k_{j_k}^{th}$ element is $ \textbf{E}[k_{j_k}] = 1$ if $f_{k_{j_k}}(\beta_{k_{j_k}}) = \beta_{k_{j_k}}$ and $\exp ( \beta_{k_{j_k}} )$ otherwise, and $\Bar{\Bar{\textbf{E}}}$ is a $\sum_{k=1}^K J_k \times \sum_{k=1}^K J_k$ diagonal matrix such that its $k_{j_k}^{th}$ diagonal element is $\Bar{\Bar{\textbf{E}}}[k_{j_k}, k_{j_k}] = 0$ if $f_{k_{j_k}}(\beta_{k_{j_k}}) = {\beta}_{k_{j_k}}$ and $\exp ( {\beta}_{k_{j_k}} )$ otherwise. 

The analytical expression of the gradient and Hessian matrix are presented in on-line Supplementary Material . Although these derivatives involve lengthy calculations as well as careful algorithmic implementation, the computational and inferential benefits of avoiding numerical approximations justify the effort. The algorithm employed for estimating the regression parameters and smoothing coefficient vector is summarized in on-line Supplementary Material . Briefly, it combines a carefully structured trust region algorithm which uses the analytical expressions of the gradient and Hessian of the log-likelihood and properly chosen starting values with a general automatic multiple smoothing parameter selection algorithm based on an approximate AIC measure. 

In practice, this results in an estimation algorithm which is general, modular, efficient and stable, working well even for problems which are non-concave and/or exhibit close to flat regions. We found this both through usage on real-world data as well as through the extensive simulation study conducted and reported in detail in on-line Supplementary Material . As expected, like any method, in the latter we found that model fitting failed to converge at times (i.e., did not achieve close to zero gradient and/or positive definite Hessian), however this occurred only for a small percentage of simulation replicates. This is in line with what we found with \texttt{survPen}, our main competitor. Further details on this can also be found in on-line Supplementary Material .

To obtain confidence intervals, we follow \cite{Wo2016} and employ the Bayesian large sample approximation $\bbeta \stackrel{\cdot}{\sim} \mathcal{N}(\widehat{\bbeta},\V_{\bbeta})$, where $\V_{\bbeta} = -\bm{H}_p(\widehat{\bbeta})^{-1}$; using $\V_{\bbeta}$ gives close to across-the-function frequentist coverage probabilities because it accounts for both sampling variability and smoothing bias, a feature that is particularly relevant at finite sample sizes. Note that applying the Bayesian approach to the modeling framework discussed in this paper follows the notion that penalization in estimation implicitly assumes that wiggly models are less likely than smoother ones, which translates into the following prior specification for $\bm\beta$, $f_{\boldsymbol{\beta}} \propto \exp\left\{- \boldsymbol{\beta}\ts \mathbf{S} \boldsymbol{\beta}/2\right\}$. 

Since the evaluation of the additive predictor in (\ref{linpredv}) and the quantities that rely on it depend on $\ff(\bbeta)$, it makes sense to obtain its distribution as well. Following \cite{Pya}, we first consider the Taylor series expansion of $\ff(\bbeta)$ around $\ff(\Tilde\bbeta)$, i.e. $\ff(\bbeta) - \ff(\Tilde\bbeta) \approx \diag(\textbf{E})\big(\bbeta - \Tilde\bbeta \big)$. This shows that $\ff(\bbeta) - \ff(\Tilde\bbeta)$ is approximately a linear function of $\bbeta$. We then recall that linear functions of normally distributed random variables follow normal distributions. This implies that $\ff(\bbeta) \overset{\cdot}{\sim} \mathcal{N}(\ff(\Tilde\bbeta), \textbf{V}_{\ff(\bbeta)})$ where $\textbf{V}_{\ff(\bbeta)} = \diag(\textbf{E}) \textbf{V}_{\bbeta} \diag(\textbf{E})$. P-values for the smooth components in the model are derived by adapting the result discussed in \cite{Wood} and using $\textbf{V}_{\ff(\bbeta)}$ as covariance matrix.

Intervals for linear functions of the model's coefficients, e.g. smooth components, can then be obtained using the result just shown for $\ff(\bbeta)$. For non-linear functions of the model's coefficients, e.g. hazard functions, instead, the intervals can be conveniently obtained by posterior simulations, hence avoiding computationally expensive parametric bootstrap or frequentist approximations, for instance.

The approximation found for $\ff(\bbeta)$ also facilitates the construction of confidence intervals for the net survival curve (either associated to an individual or a sub-population). We define the (marginal) net survival function associated to a sub-population $ \x_{pop} = \{\x_{1},\dots \x_{k}\}$ as
\begin{align*}
\Bar{S}_N(t) = \dfrac{1}{k} \sum_{\x_i \in \x_{pop} } S_N(t \mid \textbf{x}_{i}),
\end{align*}
where it is assumed that $k$ is the number of individuals belonging to the sub-population of interest. For instance, $\x_{pop}$ could be the entire population or a subgroup of interest, such as a specific age group. Keeping in mind that we are interested in finding the interval for an average over multiple net survival curves, we will have to sample from the posterior distribution of this average. Finally, on-line Supplementary Material  describes the use of the \texttt{R} package \texttt{GJRM}.

The main asymptotic results related to the proposed estimator are presented below and are based on classical assumptions from the GAM and relative survival literature and refer to model regularity conditions.

\begin{theorem}
If Assumptions A1-A8 hold (see on-line Supplementary Material ) then
\begin{enumerate}
\item[(i)] $\widehat{\bbeta} \stackrel{P}{\to} \bbeta_0$ \text{and} $\vert\vert \widehat{\bbeta}  - \bbeta_0\vert\vert =O_p(n^{-\frac{1}{2}})$,
\item[(ii)] $\sqrt{n}\left(\widehat{\bbeta}  - \bbeta_0 \right) \stackrel{d}{\to} N({\bf 0}, {\bf i}^{-1}(\bbeta_0))$,
\end{enumerate} 
where $\bbeta_0$ is the true parameters vector.
\end{theorem}

\section{Simulation study}\label{simulation}

We consider twenty scenarios, resulting from the combination of four sample sizes, $n = 200$, $500$, $1000$, $5000$, and five Data Generating Processes (DGPs) of increasing difficulty, to extensively test our method's ability to capture the true generating mechanism. We will compare the performance of our method, implemented in the \texttt{R} package \texttt{GJRM}, with a state-of-the-art model in the (relative) survival setting, i.e. \cite{fauvernier:2019} and its implementation in the \texttt{R} package \texttt{survPen}. Although other penalized relative survival model implementations exist, we consider \texttt{survPen} to be an adequate benchmark as it was in turn extensively tested against competing frameworks in the reference paper and was generally found to be superior. We will then have four fitted models: \texttt{GJRM} with each of the three allowed link functions, i.e. PH, PO and probit, and \texttt{survPen}.

As we are in a relative survival setting, we will simulate the population hazard and the excess hazard separately for each individual. The former is simulated from a piece-wise exponential distribution based on life tables from the general English population. The latter using increasingly complex functional forms with parameters set to result in approximately 40\% censoring. This level was chosen to reflect the 44.8\% censoring found in the case studies on which the simulated ones are based. For each scenario we simulate $1000$ datasets, which include also age at diagnosis and level of deprivation defined on a discrete scale between 1 (least deprived) and 5 (most deprived). 

Our method performs consistently well throughout the scenarios and over a range of metrics, with greater uncertainty generally found at the smaller sample size, as expected. In the following section, we present the results for one of the most challenging DGPs. For more details and the full set of results, we refer the reader to on-line Supplementary Material .

\subsection{General Hazards model with non-linear effect of age}\label{simulation-example}
We consider a General Hazards model, as defined in \cite{rubio:2019S}. For the $j^{th}$ observation this is given by 
\begin{align*}
    h_E^{GH}(t; \x_j) = h_0\big( t \exp[ \alpha \cdot f(\text{agec}_j) ]  \big) \exp(\x_j\ts  \bbeta) \quad \text{with} \quad h_0(r) = \frac{\phi(\frac{\ln r - \mu}{\sigma})}{\Phi(-\frac{\ln r - \mu}{\sigma}) \cdot r \sigma},
\end{align*}
where the baseline hazard $h_0(\cdot)$ is modeled using a log-normal distribution with parameters $(\mu, \sigma)$ and where $\phi(\cdot)$ and $\Phi(\cdot)$ represent respectively the density function and CDF of a standard normal. Further, $\x_j = [f(\text{agec}_j), \text{dep}_j]\ts$ where $f(\text{agec}_j) = 0.75 \sinh \big[ 0.5 \arcsinh [3 \hspace{3pt} \text{agec}_j ] \big]$ is a smooth function of the standardized age at diagnosis $\text{agec}_j$, chosen to ensure that the associated effect was not too small, and $\text{dep}_j$ is the level of deprivation. The associated (time-fixed) effect is denoted by $\bbeta$. A time-dependent effect is also assumed for  $f(\text{agec}_j)$ and is denoted by $\alpha$. The values used for the parameters $(\mu, \sigma, \alpha, \bbeta)$ are reported in Table 8 of on-line Supplementary Material .

In terms of model selection, using the AIC, the \texttt{GJRM} PH model was found to be the best among the \texttt{GJRM} models which were in turn found to be preferred to \texttt{survPen}. This holds with the exception of the case $n = 200$, for which \texttt{GJRM} probit is the best model in terms of AIC, both when compared with the other \texttt{GJRM} models as well as with \texttt{survPen}; see Table 10 of the on-line Supplementary Material . In the following, only the best \texttt{GJRM} model will be compared with the \texttt{survPen} model, to avoid cluttering the plots.

\begin{figure}[htb!]
\begin{minipage}[b]{\linewidth}
    \centering
    \includegraphics[scale = 0.45, angle = 270]{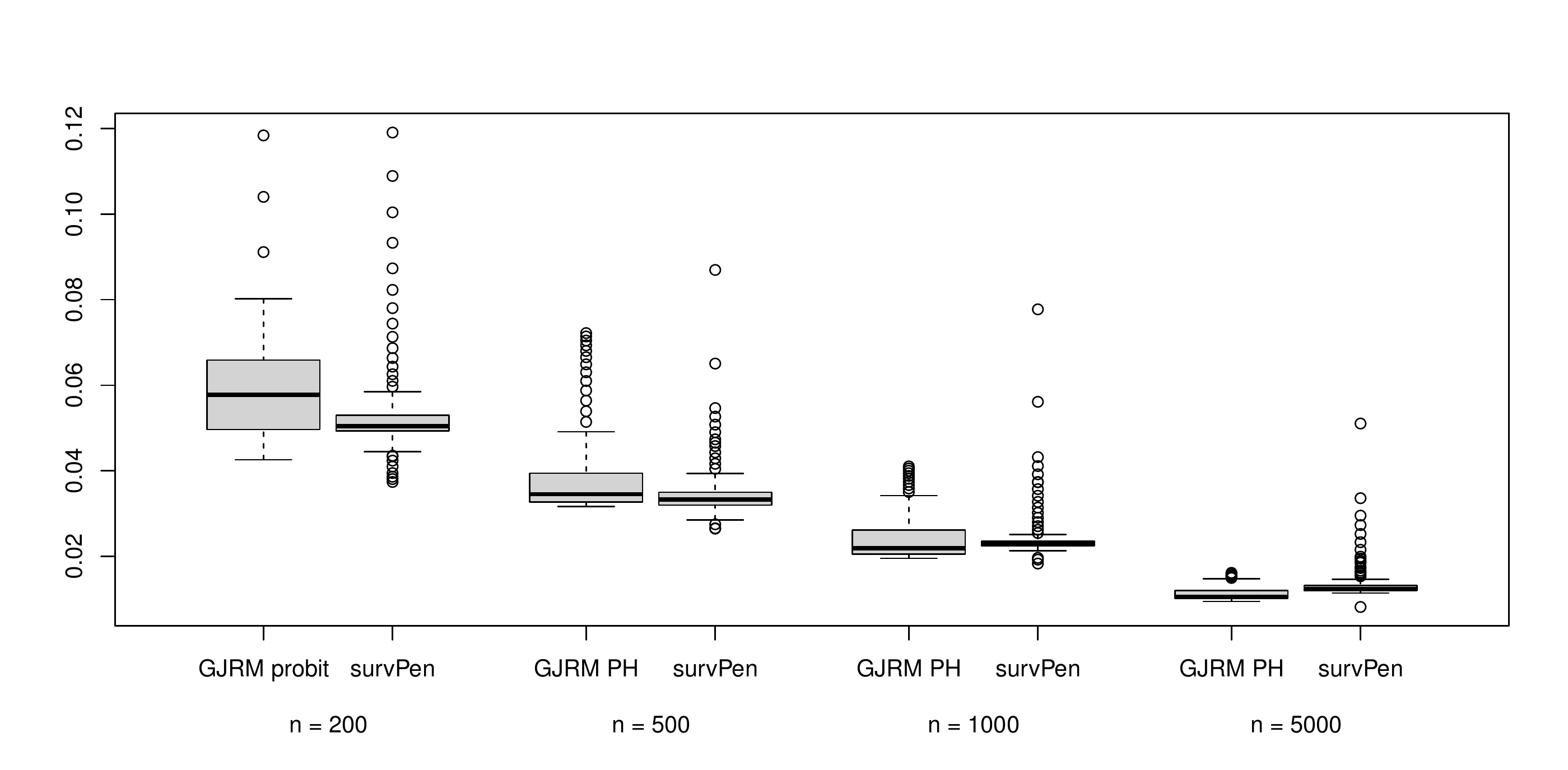}
\end{minipage}
\caption{Boxplots of RMSE of excess hazard function for each model specification under the fourth DGP.}\label{fig:RMSE5-main}
\end{figure}

\begin{figure}[htb!]
\begin{minipage}[b]{0.3\linewidth}
    \centering
    \includegraphics[scale = 0.3, angle = 270]{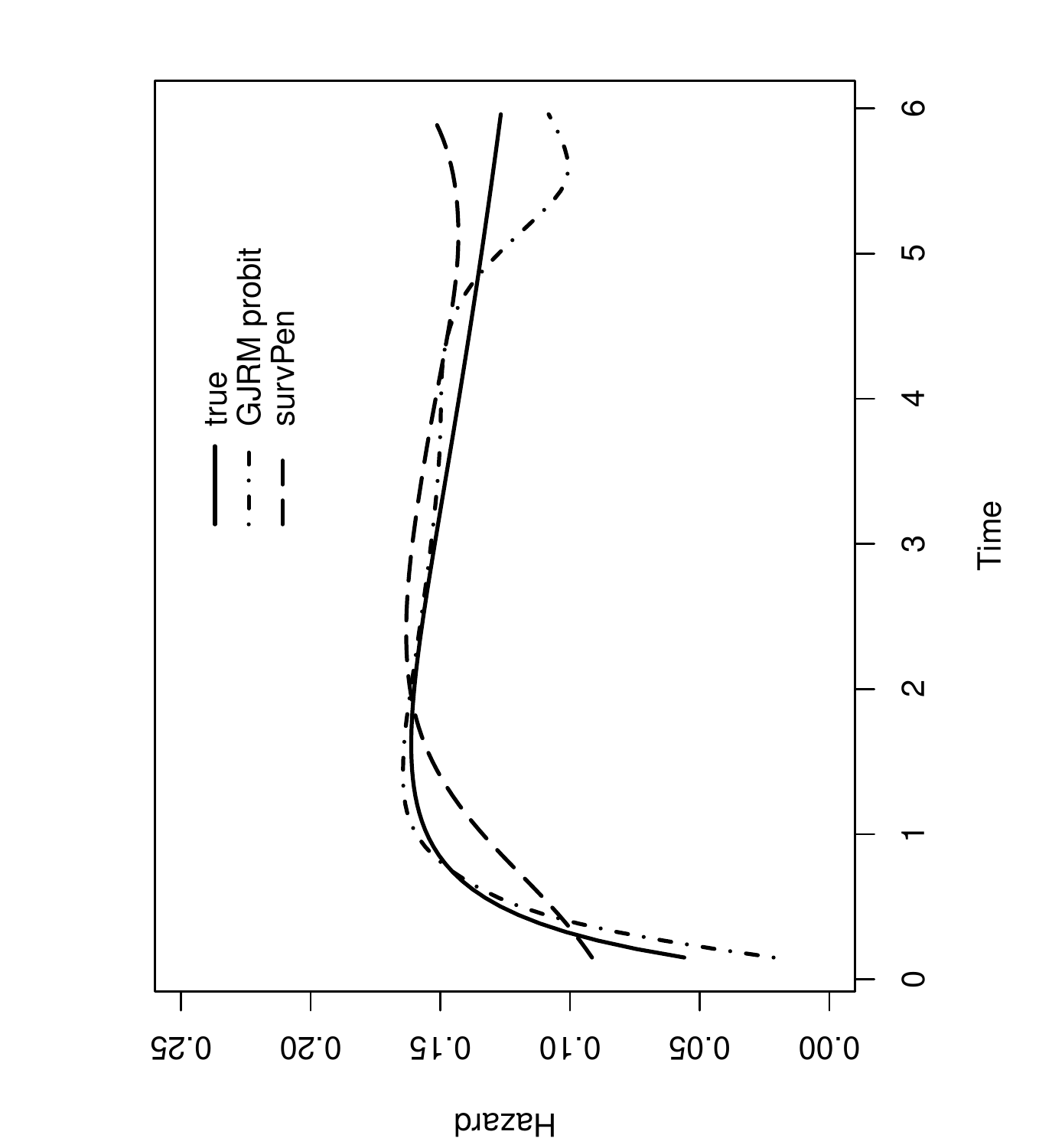}
    \subcaption{Sample size $n = 200$.}
    \label{fig:avg_hazard52.200-main}
\end{minipage}
\hfill
\begin{minipage}[b]{0.3\linewidth}
    \centering
    \includegraphics[scale = 0.3, angle = 270]{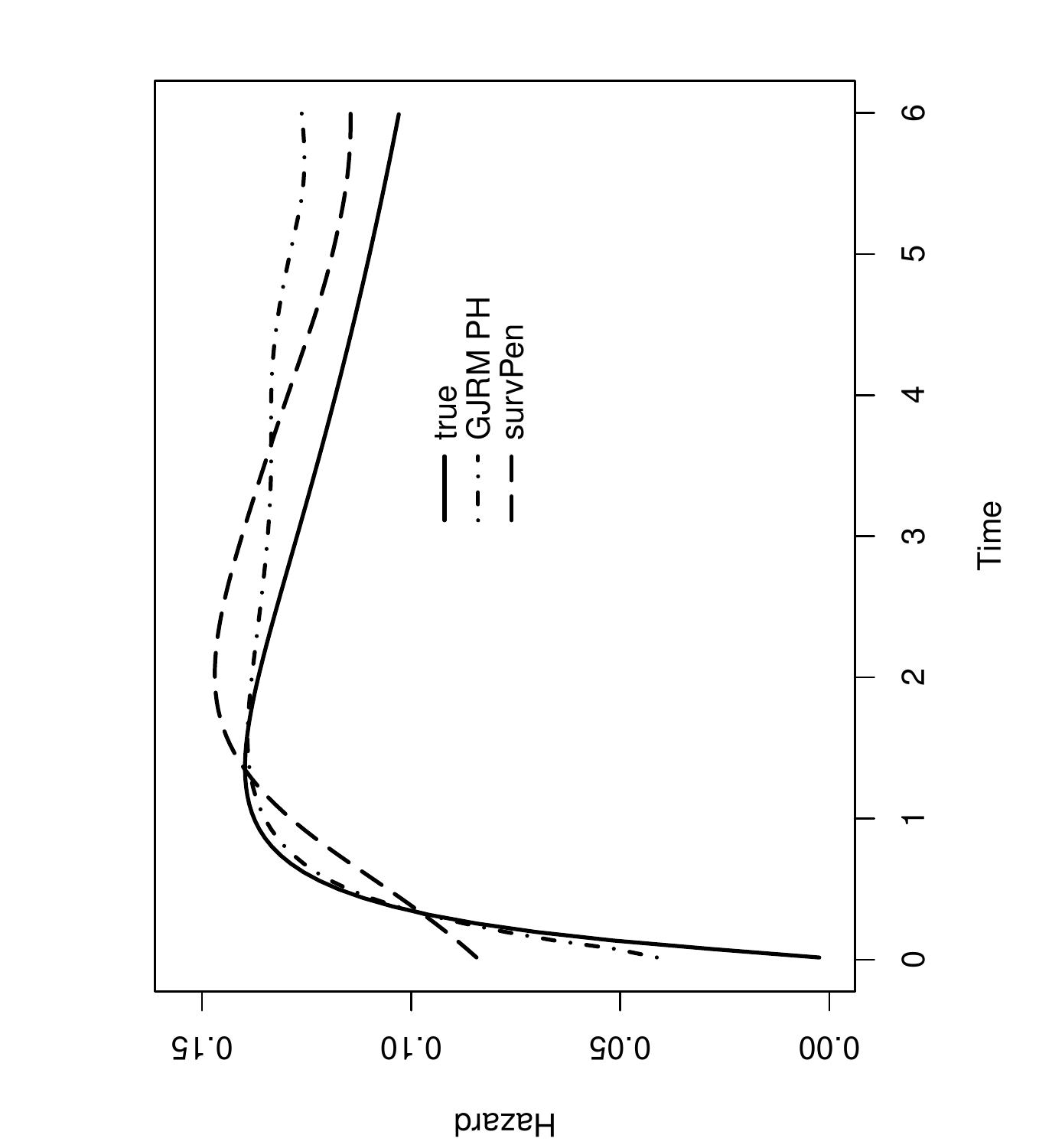}
    \subcaption{Sample size $n = 500$.}
    \label{fig:avg_hazard52.500-main}
\end{minipage}
\hfill
\begin{minipage}[b]{0.3\linewidth}
    \centering
    \includegraphics[scale = 0.3, angle = 270]{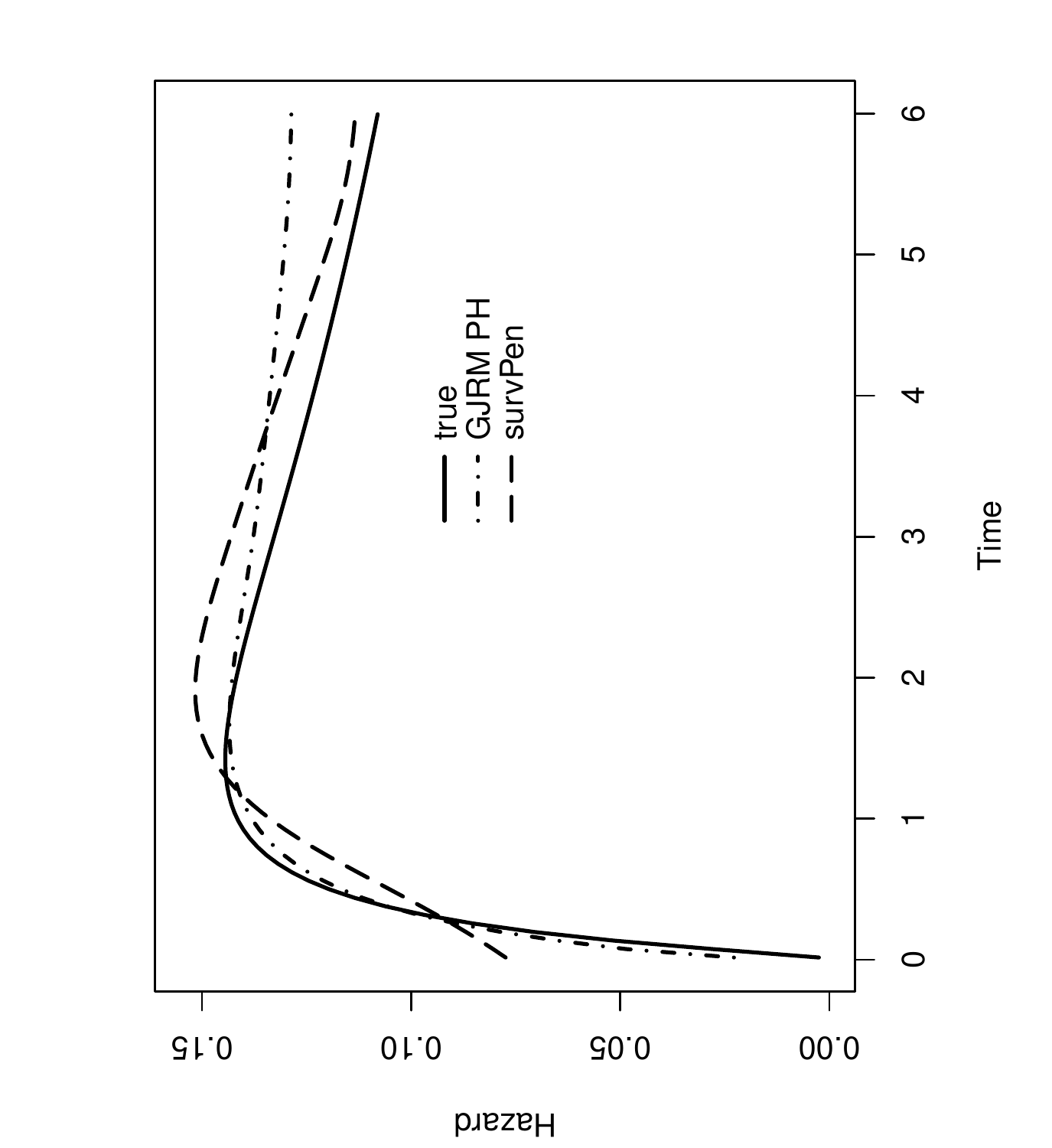}
    \subcaption{Sample size $n = 1000$.}
    \label{fig:avg_hazard52}
\end{minipage}
\vfill
\begin{minipage}[b]{\linewidth}
    \centering
    \includegraphics[scale = 0.3, angle = 270]{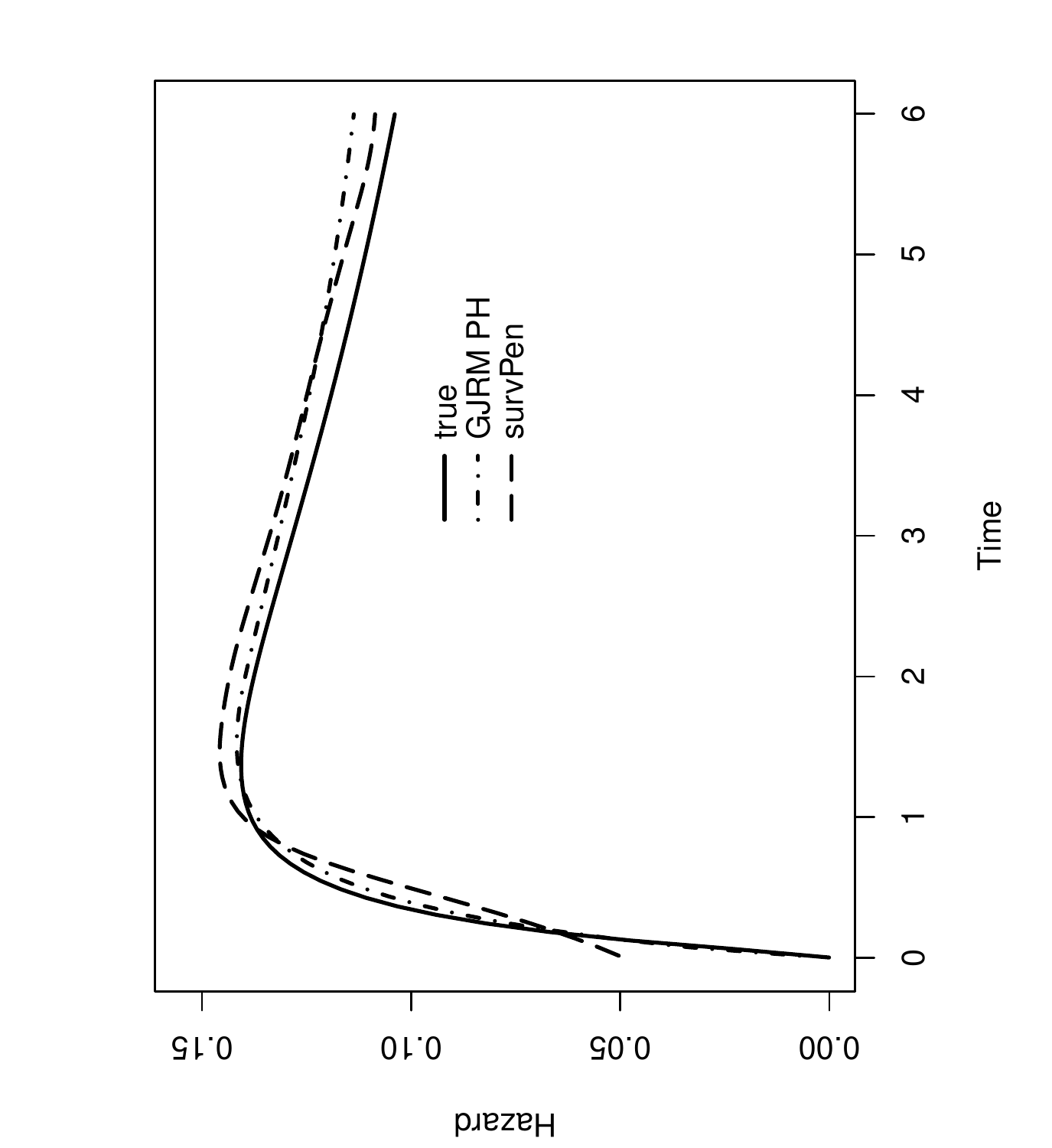}
    \subcaption{Sample size $n = 5000$.}
    \label{fig:avg_hazard52.5000}
\end{minipage}
\caption{Average estimated excess hazard for \texttt{GJRM} and \texttt{survPen} versus the true excess hazard function under the fourth DGP (the mean was taken across the excess hazard estimated on each simulated dataset).}\label{fig:avg_hazard5-main}
\end{figure}

In Figure \ref{fig:RMSE5-main} we report the boxplot of the Root Mean Square Error (RMSE) for the excess hazard for each of the two approaches and for each sample size. The boxplot of the bias is very similar so it is omitted here due to space constraints but it is reported in Figure 22 of on-line Supplementary Material . We find that the RMSE decreases as the sample size increases and that it is overall smaller but more variable for \texttt{GJRM} PH than for \texttt{survPen} at $n = 1000$ and $n = 5000$. At $n = 200$ and $n = 500$, \texttt{survPen} outperforms \texttt{GJRM} probit and \texttt{GJRM} PH respectively; from Figure \ref{fig:avg_hazard5-main}, we see that the \texttt{survPen} estimated excess hazard curve is not very close to the true curve although it is overall better than the one produced by \texttt{GJRM}. In general, \texttt{GJRM} mostly leads to curves which trace the true excess hazard more closely but, when they do not, specifically in the middle and final times, they contribute to higher overall values of RMSE. We find this behavior in the average estimated excess hazard plots reported in Figure \ref{fig:avg_hazard5-main}, where \texttt{GJRM} captures the first portion of the true excess hazard relatively well, even at the lowest sample sizes, but departs from it in the final times. \texttt{survPen}, instead, struggles to capture the initial steeply increasing portion of the true excess hazard while it is closer to the true curve in the final times. At the highest sample size, \texttt{GJRM} improves greatly its fit also in the final times, becoming almost indistinguishable from the true curve across the entire time range. \texttt{survPen} also greatly improves its fit to the first portion of the true curve but does not match \texttt{GJRM}.

We omit the analysis of the population net survival as both methods perform very well, with almost perfectly overlapping estimated average curves when compared with the true survival. Consequently the bias and RMSE are markedly smaller than for the estimated excess hazard with the two methods overall comparable in magnitude and variability. These details can nonetheless be found in on-line Supplementary Material .

In conclusion, the simulation study shows that \texttt{GJRM} and \texttt{survPen} are comparable and both able to adequately capture even complex DGPs. In particular, \texttt{survPen} seems to perform better when \textit{n} is smaller and, while capturing satisfactorily the overall shape of the true excess hazard, it appears slightly shifted and flatter. \texttt{GJRM} performs less well at the lowest sample size but the fit improves dramatically as the sample size increases. As all of the datasets considered in the applications are characterized by a number of patients which is well above the largest sample size considered here, empirical performance is not likely to be a matter of concern. Finally, both approaches are characterized by similar model specifications and run times for model fitting, however \texttt{survPen} is slower when calculating (sub-)population net survival estimates. For more details on these aspects and for the full set of results we refer the reader to on-line Supplementary Material .

\section{Case Studies}\label{sec:applications}

Cancer research strives to provide an accurate picture of the evolving cancer burden, as well as documenting existing inequalities, using a variety of key indicators, including cancer survival. In this section, we present two case studies which aim at investigating inequalities in net survival for patients diagnosed in England.
The first case study aims at investigating socio-economical inequalities in net survival for the top three incident cancer types (breast, colon and lung), but which have differential levels of survival as confirmed by previous research \citep{quaresma:2015}, and the second case study aims at studying geographical disparities in net survival for colon cancer patients in England. 

For these case studies, individual cancer records were obtained from the National Cancer Registry at the Office for National Statistics (ONS) on all adult patients (aged 15-99 years) diagnosed with a first, primary, invasive malignancy of the breast (women), colon and lung during 2010 in England. All cancer records were followed-up by the National Health Service Central Register, who routinely update these records with information about each patient's vital status. These cases were followed up until the 31st December 2015.  Survival times were measured from the date of diagnosis until the date of death or last time of follow-up.
The individual patient and tumour-specific variables available for this analysis were: full date of diagnosis, last follow-up and death times, vital-status indicator (which takes value 1 if the patient died of the cancer of interest and 0 otherwise), age at diagnosis (recorded as a continuous variable) and deprivation category (1-least deprived to 5-most deprived) defined according to the quintiles of the distribution of the Income Domain scores of the 2011 England Indices of Multiple Deprivation, NHS England Regions and Local Offices of residence (\href{https://geoportal.statistics.gov.uk/datasets/nhs-england-region-local-office-april-2019-en-bgc}{14 geographical regions}), and tumor stage at diagnosis (I-IV). 
Background mortality rates were obtained for each cancer patient from population life tables for England defined for each calendar year in 2010-2015, and stratified by single year, age, sex, deprivation category and Government Office Region of residence. 

Additional examples of data analyses, including of spatial effects on cancer survival, can be found on the public repository {\small\texttt{https://github.com/FJRubio67/LBANS/}}, where we consider the \href{https://simulacrum.healthdatainsight.org.uk/}{Simulacrum dataset} and the \texttt{LeukSurv} dataset from the \texttt{R} package \texttt{spBayesSurv}.

\subsection{Socio-demographic inequalities in breast, colon and lung cancer survival in England}\label{appl1}

In this case study, we compare the net survival curves for the most deprived and least deprived groups of the population for the three major cancer types.
The breast cancer in women dataset contains $n= 38,636$ complete cases, with median age $62.8$ years, from which $n_o=9,169$ patients died within the follow-up period (76.2\% censoring). 
The colon cancer in men dataset contains $n=11,106$ complete cases, with median age $72.7$ years, from which $n_o=6,126$ patients died within the follow-up period (44.8\% censoring). 
The colon cancer in women dataset contains $n=9,999$ complete cases, with median age $74.8$ years, from which $n_o=5,520$ patients died within the follow-up period (44.8\% censoring). 
The lung cancer in men dataset contains $n=18,609$ complete cases, with median age $72.8$ years, from which $n_o= 17,286$ patients died within the follow-up period (7.1\% censoring). 
The lung cancer in women dataset contains $n=14,920$ complete cases, with median age $73.2$ years, from which $n_o=13,418$ patients died within the follow-up period (10.1\% censoring). 
We chose to analyse these datasets as the cancers they involve are the three most commonly diagnosed types in England \citep{ONS:2011} with each having differential levels of survival. Previous research investigating trends in cancer survival in England since the 1970s has, in fact, identified three broad groups of cancers based on their levels of survival: those with good prognosis, including breast cancer, those with moderate survival levels, including colon cancer, and those with very poor prognosis, including lung cancer, for which little improvement has occurred in the past 40 years up to 2010 \citep{quaresma:2015}.

Table \ref{tab:GJRM_survivals} shows the net survival estimates and the corresponding $95\%$ confidence intervals, at $1,3$ and $5$ years after diagnosis, using \texttt{GJRM}. Models equivalent to those specified for \texttt{GJRM} were fitted using \texttt{survPen} as well; the AIC favored the \texttt{GJRM} models in all cases. The net survival estimates obtained using \texttt{survPen} were very close to those obtained with the best \texttt{GJRM} model; they can be found in Table 13 of on-line Supplementary Material . 

Note, in particular, that the results reported in Table \ref{tab:GJRM_survivals} correspond to the best model selected using the AIC among nine models obtained by combining the three different allowed links (PH, PO and probit) with three different definitions of the additive predictor. The first of these specifications includes a linear effect of age at diagnosis and takes the form $\eta_i = \beta_0 + \text{dep}_i\ts \bbeta_1 + \text{agec}_i \beta_2 + s_1(\log(t_i))$, the second includes a non-linear effect of age at diagnosis and takes the form $\eta_i = \beta_0 + \text{dep}_i\ts \bbeta_1 + s_1(\log(t_i)) + s_2(\text{agec}_i)$ while the third one includes a non-linear and time-dependent effect of age at diagnosis and takes the form $\eta_i = \beta_0 + \text{dep}_i\ts  \bbeta_1  + s_1(\log(t_i)) +  s_2(\text{agec}_i) + s_3(\log(t_i), \text{agec}_i)$.
Here $\text{dep}_i$ represents the level of deprivation, defined on a discrete scale from 1 (least deprived) to 5 (most deprived), $s_1(\cdot)$ a monotonic P-spline taken over the logarithm of time chosen to ensure the monotonicity of the survival function as explained in Section \ref{pll}, $s_2(\cdot)$ a cubic regression spline taken over the standardized age at diagnosis of cancer $\text{agec}_i$ and $s_3(\cdot)$ a pure tensor product interaction between standardized age at diagnosis and time, whose marginals are also cubic regression splines. This is how time-dependent effects are included in the models specified in the applications and in the simulation study. Note that the term ``pure'' refers to the fact that sum-to-zero constraints remove the unit function from the span of the marginals, with the result that the tensor product basis will not include the main effects. These, in fact, would result from the product of a marginal basis with the unit functions in the other marginal bases. In other terms, this specification enables us to model the main effects and the interaction term separately, thus leading to more flexibility as the main effects are allowed to have different complexity from their associated effects in the interaction term \citep{Wood}. With regard to the penalty associated with the non-linear term $s_2(\text{agec}_i)$, this takes the form of the quadratic penalty defined in Section \ref{ap} with $\textbf{D}_k$ given by the integrated square second derivative of the basis functions, i.e. $\int \textbf{d}_k(z_k) \textbf{d}_k(z_k)\ts  d z_k$ with the $j_k^{th}$ element of $\textbf{d}_k(z_k)$ defined as $\partial^2 b_{k j_k} (z_k) / \partial z_{k}^2$. The penalty associated with the non-linear pure interaction term $s_3(\log(t_i), \text{agec}_i)$ is, instead, more complex as it entails combining two penalties, each corresponding to one of the arguments of the smooth function. These are summed after being weighted by smoothing parameters, which thus serve the purpose of controlling the trade-off between the smoothness in each of the two directions; for more details on this we refer the reader to Chapter 5 of \cite{Wood}. The best model according to the AIC is the one obtained by combining the probit link with the last specification. This is the case for all five datasets. 

\begin{figure}[htb!]
\begin{center}
\begin{tabular}{ccc}
\includegraphics[scale= 0.3]{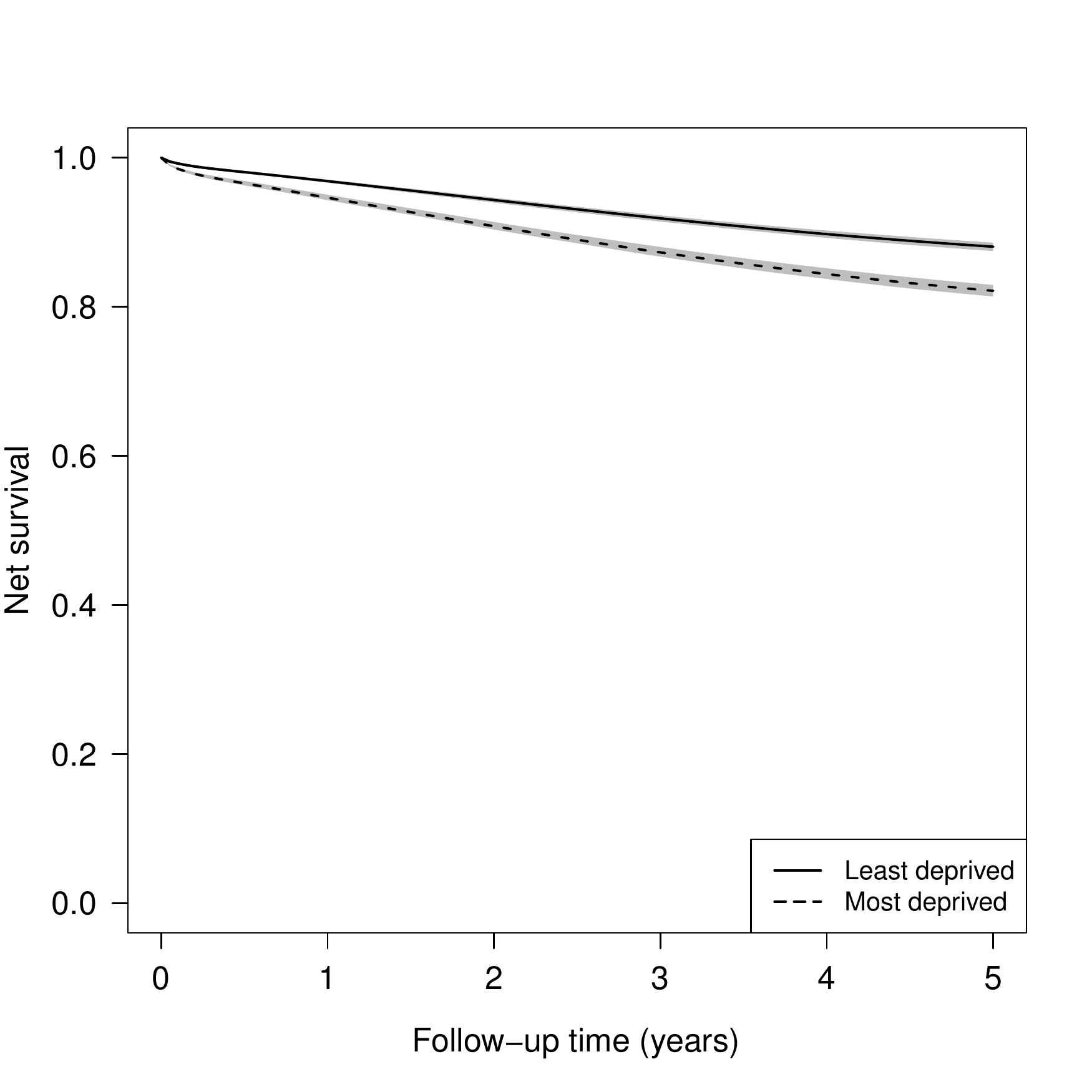} &
\includegraphics[scale= 0.3]{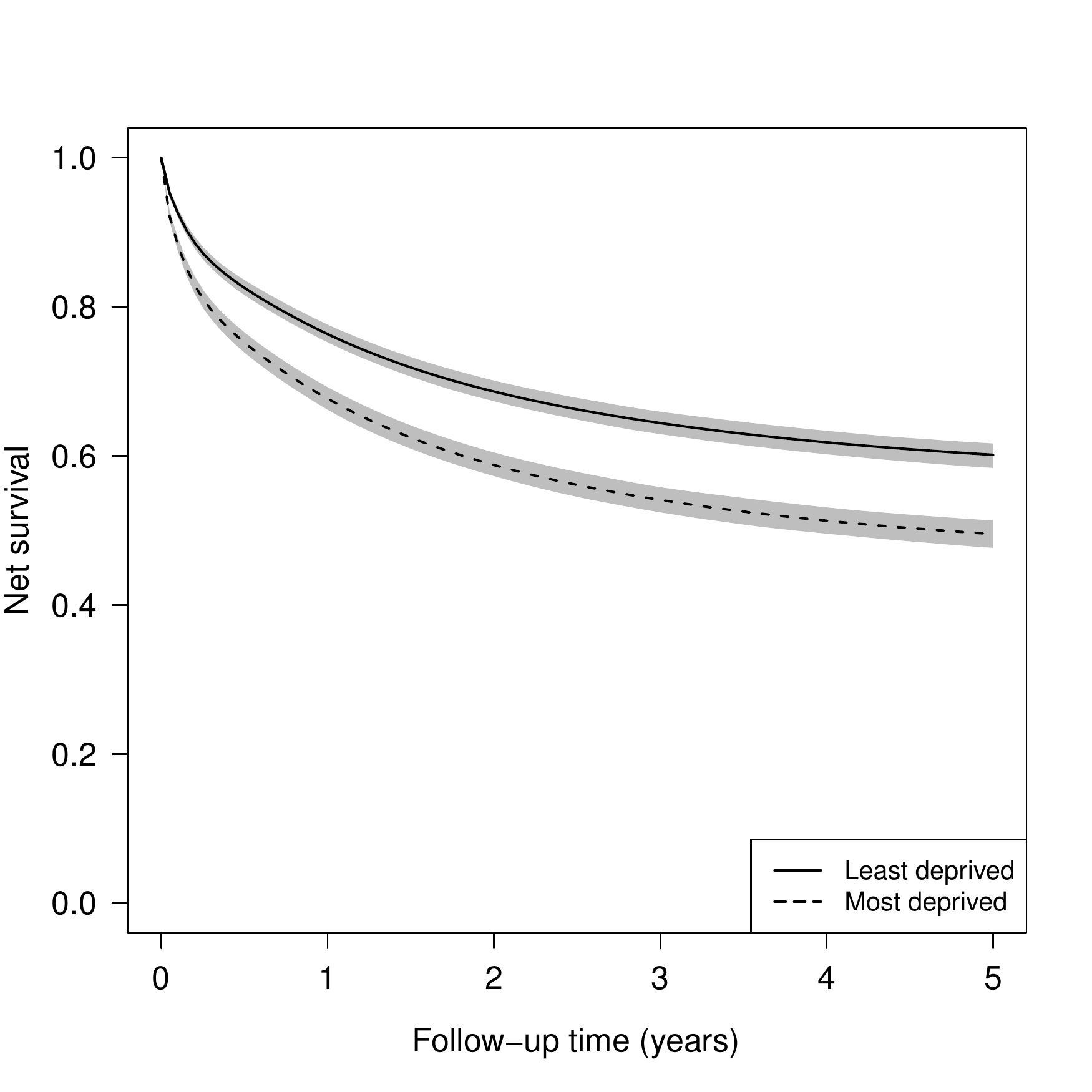} &
\includegraphics[scale= 0.3]{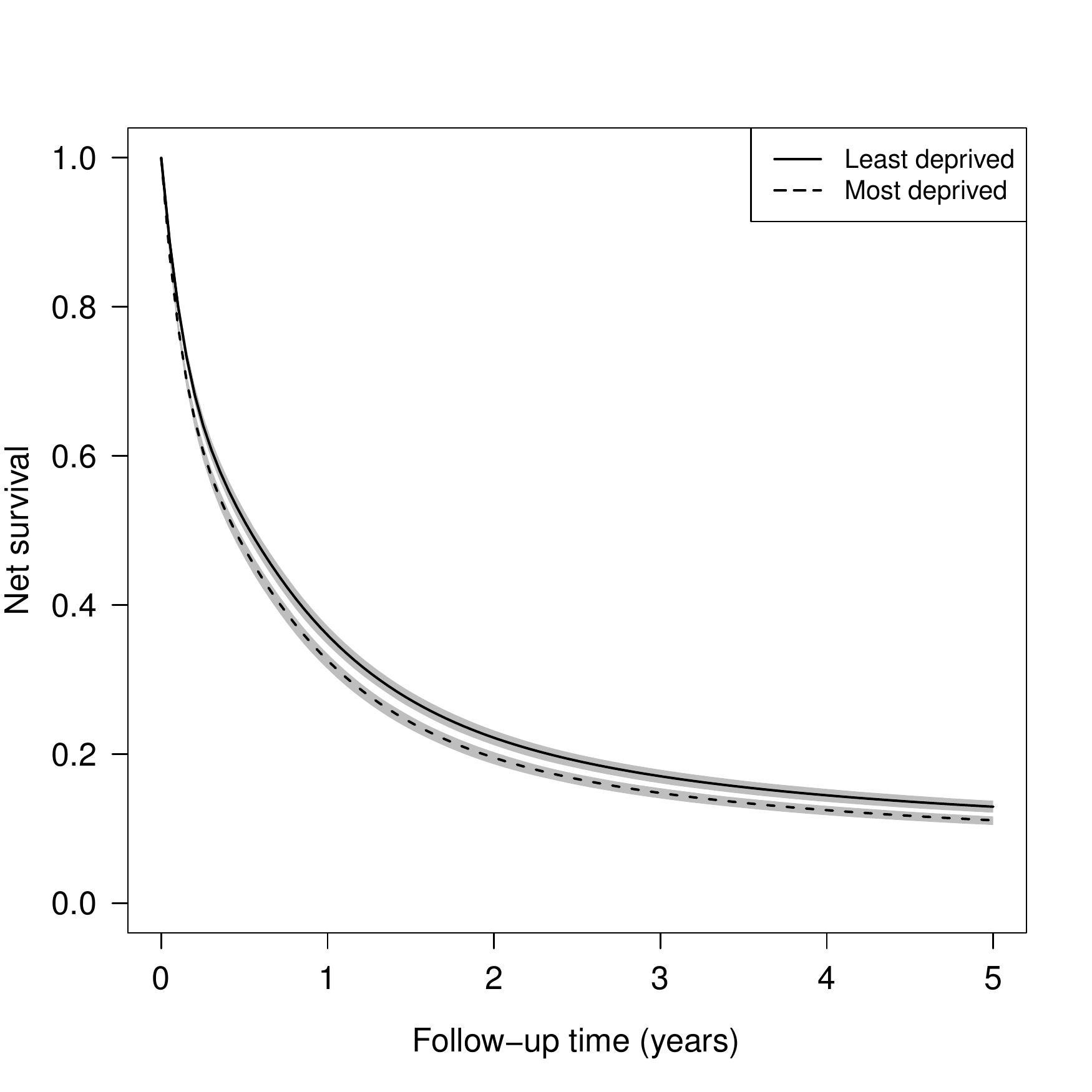} \\
(a) & (b) & (c) 
\end{tabular}
\caption{Net survival for all adult women (aged 15-99 years) diagnosed during 2010 in England (least deprived \emph{vs.}~most deprived): (a) breast cancer; (b) colon cancer; (c) lung cancer.}
\label{fig:ns3cancer}
\end{center}
\end{figure}

Figure \ref{fig:ns3cancer} presents the net survival and the population excess hazard curves for data on breast, colon and lung cancer for female patients with deprivation categories 1 and 5. For completeness, we present the output for the best model as well as the smooths of age and $\log(t_i)$ for the colon cancer in men dataset in on-line Supplementary Material . Here, further details on how the non-linear and time-dependent effects can be specified in \texttt{R} can also be found.

Very high levels of survival were observed for women diagnosed with breast cancer in 2010 (above 80\% at five years after diagnosis), moderate levels of survival for both men and women diagnosed with colon cancer (above 50\% at five years after diagnosis), and very low survival for patients diagnosed with lung cancer in both genders. For all cancers, net survival was always lower for the most deprived group of patients at all times after diagnosis (Table \ref{tab:GJRM_survivals}). From the three cancers, the largest differences between the most deprived and the most affluent groups were observed for colon cancer patients.

\begin{table}[htb!]
\centering
\begin{tabular}{c c c c}
\hline
(yrs) &  pop &  dep 1 & dep 5 \\ 
\hline
\multicolumn{4}{c}{breast cancer}
\\
 1 & 0.96 (0.96, 0.96) & 0.97 (0.97, 0.97) & 0.95 (0.94, 0.95) \\ 
 3 & 0.90 (0.89, 0.90) & 0.92 (0.91, 0.92) & 0.87 (0.87, 0.88) \\ 
 5 & 0.85 (0.85, 0.86) & 0.88 (0.87, 0.89) & 0.82 (0.81, 0.83) \\ 
\multicolumn{4}{c}{colon cancer (men)}
\\
1 & 0.74 (0.74,  0.75) & 0.77 (0.76, 0.78) & 0.71 (0.69, 0.72) \\ 
3 & 0.62 (0.61, 0.62) & 0.65 (0.63, 0.66) & 0.57 (0.55, 0.59) \\ 
5 & 0.57 (0.56, 0.58) & 0.60 (0.58, 0.62) & 0.52 (0.50, 0.54) \\ 
\multicolumn{4}{c}{colon cancer (women)}
\\
 1 & 0.72 (0.71, 0.73) & 0.76 (0.75, 0.78) & 0.68 (0.66, 0.69) \\ 
 3 & 0.60 (0.58, 0.60) & 0.64 (0.63, 0.66) & 0.54 (0.52, 0.56) \\ 
 5 & 0.55 (0.54, 0.56) & 0.60 (0.59, 0.62) & 0.50 (0.47, 0.52) \\ 
 \multicolumn{4}{c}{lung cancer (men)} 
\\
1 & 0.30 (0.29, 0.30) & 0.31 (0.30, 0.32) & 0.29 (0.28, 0.30) \\ 
3 & 0.13 (0.12, 0.13) & 0.13 (0.13, 0.14) & 0.12 (0.12, 0.13) \\ 
5 & 0.09 (0.09, 0.09) & 0.10 (0.09, 0.10) & 0.09 (0.08, 0.09) \\ 
\multicolumn{4}{c}{lung cancer (women)}
\\
1 & 0.34 (0.34, 0.35) & 0.36 (0.35, 0.37) & 0.33 (0.32, 0.34) \\ 
3 & 0.16 (0.15, 0.16) & 0.17 (0.16, 0.18) & 0.15 (0.14, 0.15) \\ 
5 & 0.12 (0.12, 0.12) & 0.13 (0.12, 0.14) & 0.11 (0.11, 0.12) \\ 
\hline
\end{tabular}
\caption{Net survival at 1, 3 and 5 years after diagnosis (``yrs'') with 95\% confidence interval between brackets for all adult (aged 15-99 years) women diagnosed with breast cancer and men and women diagnosed with colon and lung cancer during 2010 in England: population net survival (``pop''), net survival for the least deprived patients (``dep 1'') and net survival for the most deprived patients (``dep 5''). The estimates were obtained using the \texttt{R} package \texttt{GJRM}.} 
\label{tab:GJRM_survivals}
\end{table}

\subsection{Modeling spatial effects on colon cancer survival in England}\label{appl2}
For this second case study, we highlight how to incorporate spatial effects in our methodology to analyze geographical inequalities in net survival.
The dataset contains $n=9,379$ complete cases, with median age $72.08$ years, from which $n_o=4,859$ patients died within the follow-up period (48.2\% censoring). There were $2,092$ patients with deprivation level 1 (least deprived), $2,126$ with deprivation level 2, $1,946$  with deprivation level 3, $1,748$ with deprivation 4, and $1,467$ with deprivation level 5 (most deprived). Among all patients, $1,316$ were diagnosed with stage 1 tumor, $2,880$ with stage 2, $2,779$ with stage 3, and $2,404$ with stage 4.
We use as the geographical unit of analysis the NHS England Regions and Local Offices of residence. 

We fitted the same nine models described in Section \ref{appl1} with the only difference being that the additive predictors now include a spatial variation term as well. The best model in terms of AIC is that obtained by combining a PH link function with the most complex additive predictor specification, i.e. $\eta_i = \beta_0 + \text{dep}_i\ts  \bbeta_1 + s_1(\log(t_i)) + s_2(\text{agec}_i) + s_3(\log(t_i), \text{agec}_i) + s_{\text{spatial}}(\text{region}_i)$, where $s_{\text{spatial}}(\cdot)$ models the English National Health Service (NHS) regions and local offices of residence for individual \textit{i}, indicated by $\text{region}_i$, using an MRF approach. In practice, considering \textit{R} distinct regions, (\ref{bfunc}) takes the form $s_{\text{spatial}}(\text{region}_i) = \bbeta_k\ts \boldsymbol\delta_i^{(reg)}$ where $\bbeta_k = [\beta_{k 1}, \dots, \beta_{k R}]\ts $ is the vector of effects associated to each region and $\boldsymbol\delta_i^{(reg)} = [\delta_{i 1}^{(reg)}, \dots, \delta_{i R}^{(reg)}]\ts $ is such that $\delta_{i r}^{(reg)} = 1$ if individual \textit{i} belongs to region \textit{r} and 0 otherwise, for every $i = 1, \dots, n$ and $r = 1, \dots, R$. To ensure that neighboring regions have similar effects $\beta_{k r}$, we penalize the sum of squared differences between $\beta_{k r}$ values for all pairs of neighboring regions. In other terms, we impose the penalty
\begin{align*}
    \text{Pen}(\boldsymbol\bbeta_k) = \sum\limits_{r = 1}^m \sum\limits_{\substack{ q \in \text{nei}(r) \\ q > r  }} (\beta_{k r} - \beta_{k q})^2,
\end{align*}
where taking only the terms $q > r$ in the inmost summation ensures that the squared difference between a given pair is taken only once. Further, $\text{nei}(r)$ represents the set of neighbors for region \textit{r}. This can be re-written in terms of the quadratic penalty introduced above by defining an $R \times R$ matrix $\textbf{D}_k$ with diagonal elements $n_r$ given by the number of neighbors for region \textit{r} and off-diagonal elements $\textbf{D}_{k}[q,r] = -1$ if $q \in \text{nei}(r)$ and 0 otherwise, for $q,r = 1, \dots, R$. Note that the penalty can also be viewed as being induced by an improper Gaussian prior $\boldsymbol\gamma \sim \mathcal{N}(\textbf{0}, \tau \textbf{D}_k^{-})$, where $\tau$ is some precision parameter which replaces the smoothing parameter $\lambda_k$ from the penalized likelihood framework through the equality $\tau = \lambda_k^{-1}$. The $\boldsymbol\gamma$ and the neighborhood structure can then be viewed as an intrinsic Gaussian MRF with precision matrix $\textbf{D}_k$ \citep{Rue05}.

The setup of the spatial effects in \texttt{R} is straightforward. The region boundaries, in fact, are openly accessible on the \href{https://geoportal.statistics.gov.uk/}{Office for National Statistics website}. The regions can then be setup using the \texttt{GJRM} function \texttt{polys.setup()} and used in the model specification as the argument of the MRF smooth. For further details on how the models have been specified using the \texttt{R} package \texttt{GJRM} we refer the reader to on-line Supplementary Material .

\begin{figure}[htb!]
\centering
\begin{tabular}{c c}
\includegraphics[scale=0.4]{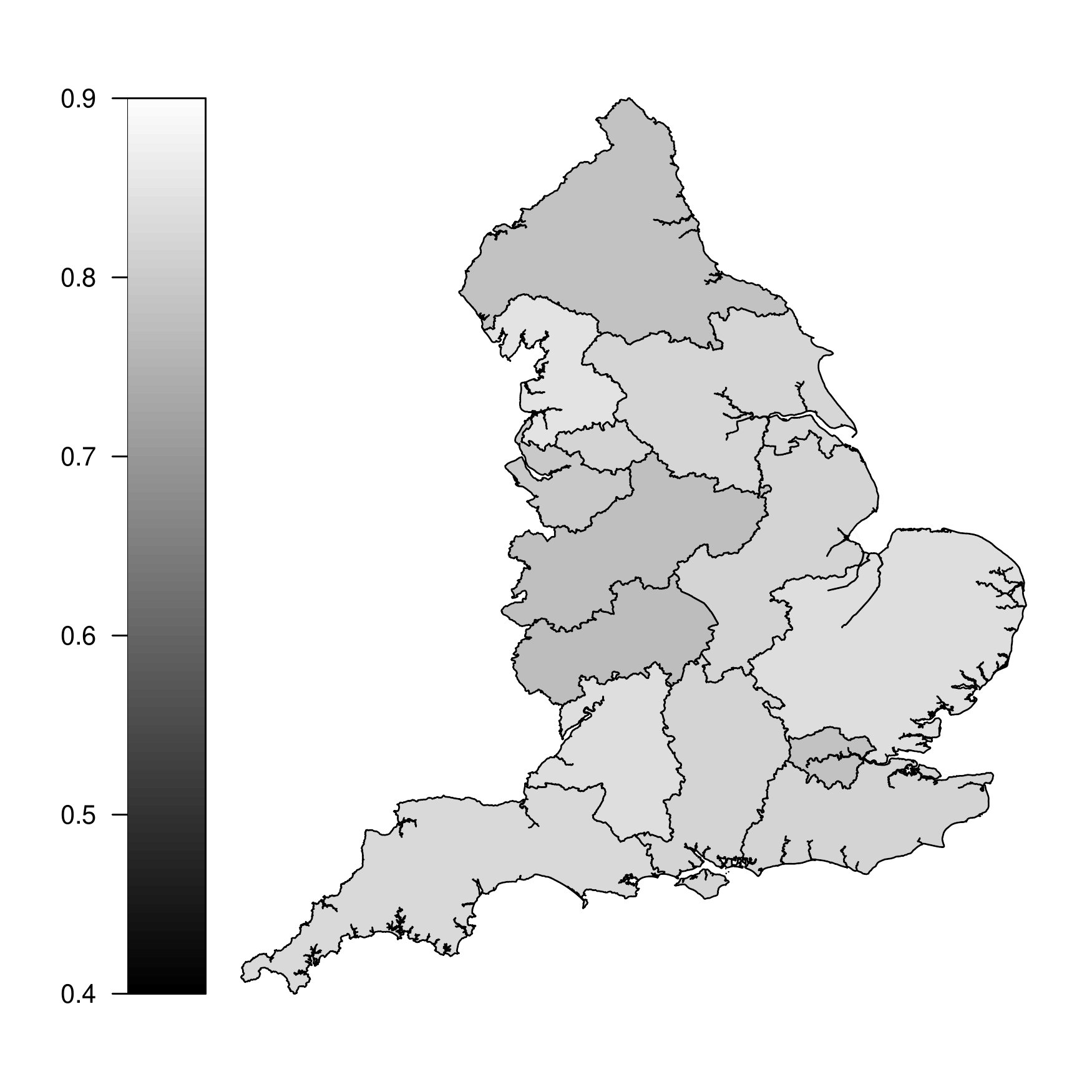}  &
\includegraphics[scale=0.4]{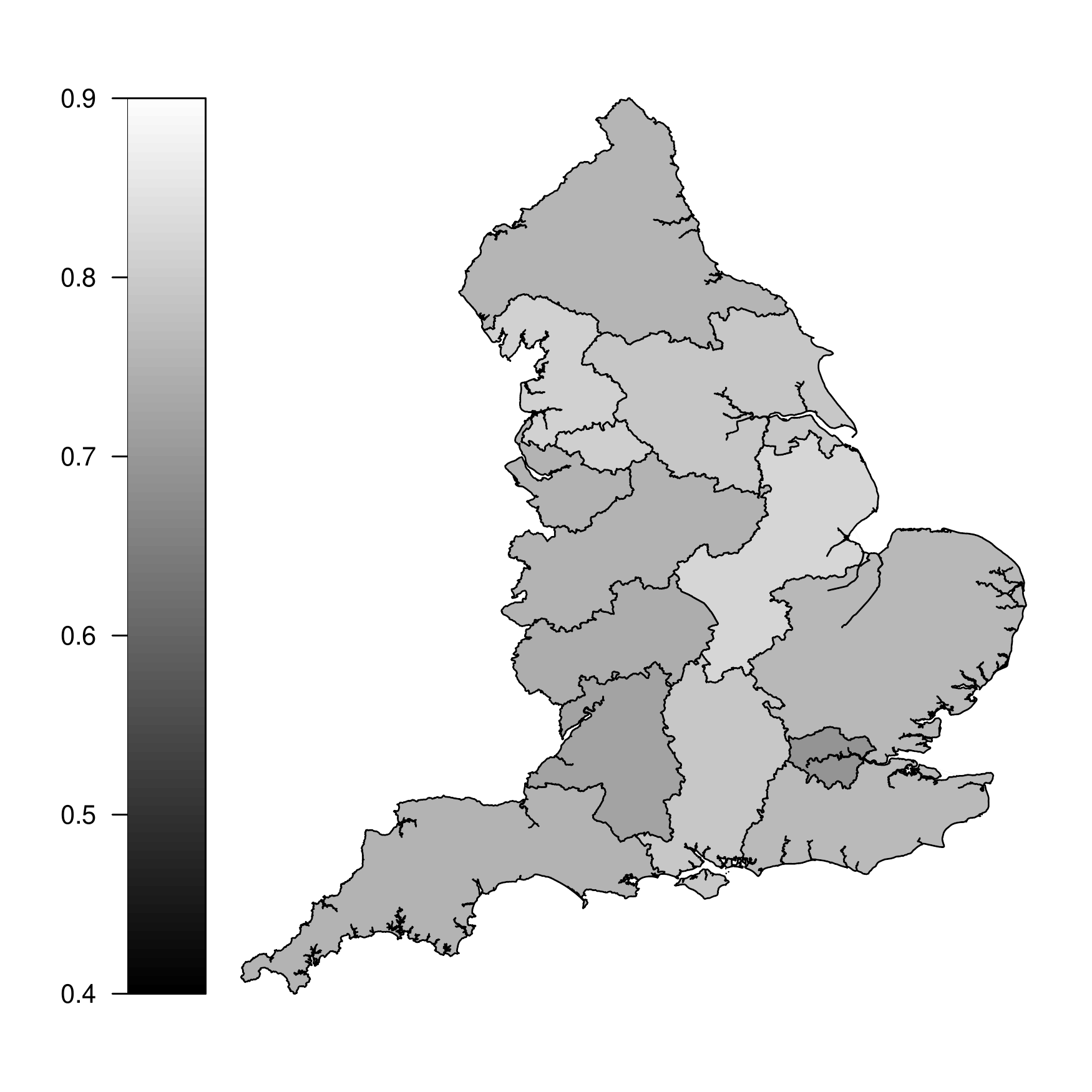}  \\
(a) & (b) \\
\includegraphics[scale=0.4]{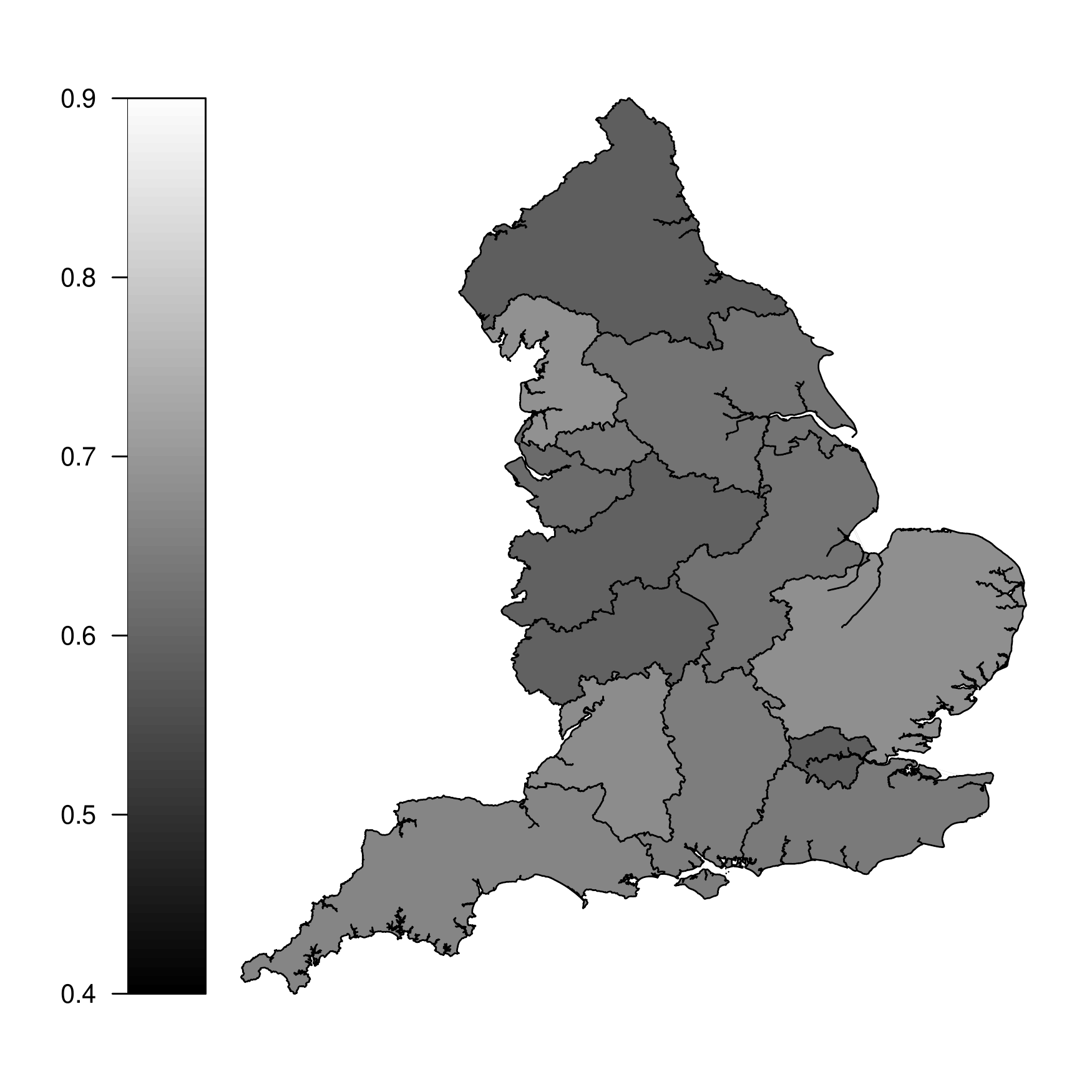}  &
\includegraphics[scale=0.4]{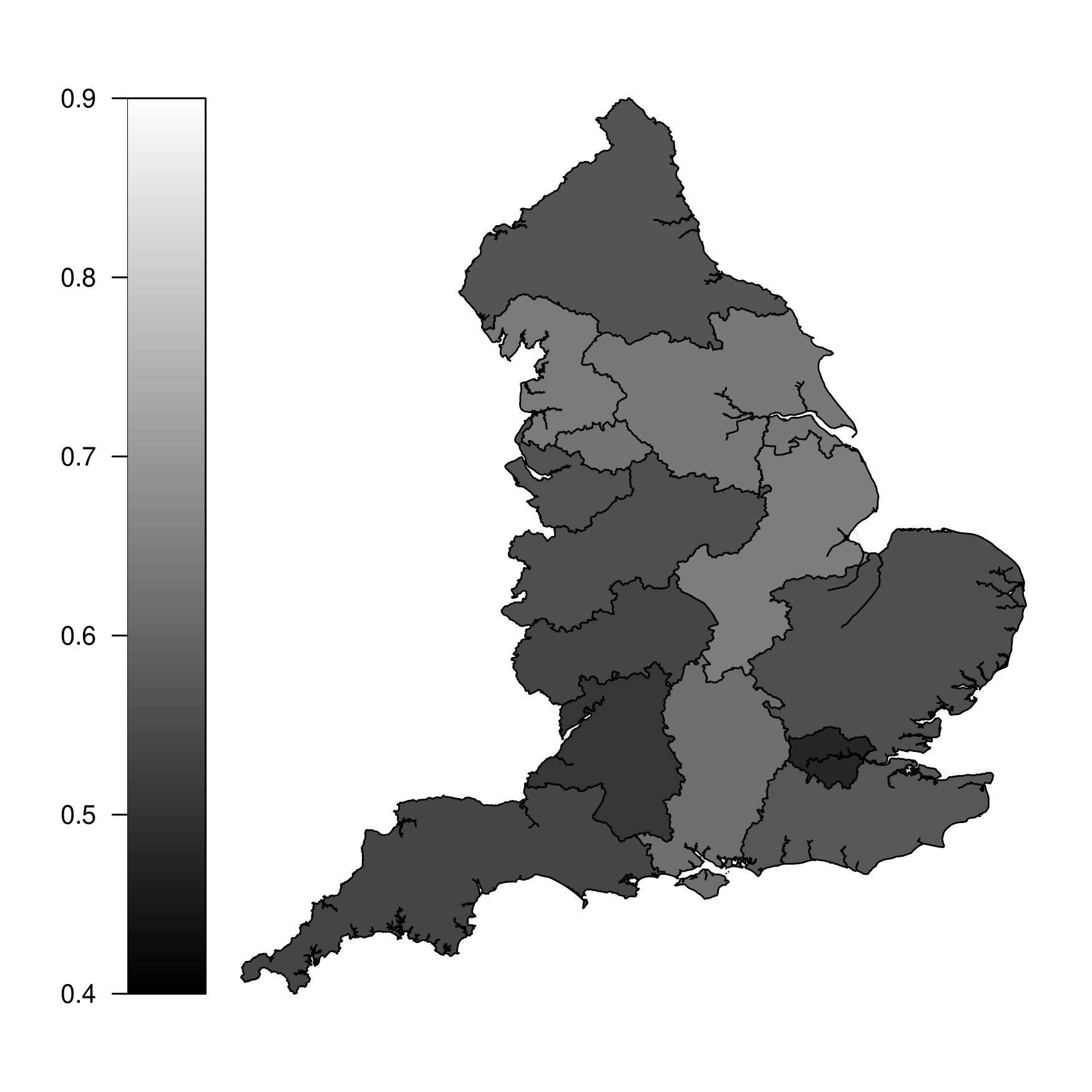}  \\
(c) & (d) 
\end{tabular}
\caption{Maps of net survival for all adult male patients (aged 15-99 years) diagnosed with colon cancer during 2010 in England: (a) 1-year net survival (least deprived), (b) 1-year net survival (most deprived), (c) 5-year net survival (least deprived), (d) 5-year net survival (most deprived).}
\label{fig:mapsNS}
\end{figure}

We report net survival at 1 and 5 years after diagnosis in Figure \ref{fig:mapsNS}: Figures \ref{fig:mapsNS}a and \ref{fig:mapsNS}b present the results for the least deprived patients, and Figures \ref{fig:mapsNS}c and \ref{fig:mapsNS}d for most deprived patients. These plots have been obtained using the \texttt{GJRM} function \texttt{polys.map()}. In line with the results presented in the first case study, we observe that net survival is consistently lower for the most deprived category across all regions, which becomes more evident at 5 years after diagnosis. Moreover, we notice some variability in net survival by region. Public Health England annually reports an Index of Cancer Survival \citep{quaresma:2015}, where, in previous years, a clear north-south gradient in survival was reported (\href{https://www.gov.uk/government/publications/cancer-survival-index-for-clinical-commissioning-groups/index-of-cancer-survival-for-clinical-commissioning-groups-in-england-adults-diagnosed-2002-to-2017-and-followed-up-to-2018}{Index of cancer survival for Clinical Commissioning Groups in England}). However, this gradient has been consistently narrowing, which is also in line with the results reported in Figure \ref{fig:mapsNS}. This type of targeted descriptive spatial summaries are crucial for generating hypotheses which may  serve as a basis for conducting more in-depth investigations about the factors driving the observed inequalities.

\section{Discussion}\label{discussion}

We have proposed a unifying framework for excess hazard estimation using a link-based additive model formulation, which allows for a variety of link functions, several types of covariate effects, and for all types of censoring and left-truncation. Estimation is based on a carefully constructed efficient and stable penalized likelihood-based algorithm. Under standard conditions for generalized additive models adapted to the excess hazard setting, we have shown consistency and asymptotic normality of the estimators. An intuitive implementation of the proposed methodology is available in the \texttt{R} package \texttt{GJRM}, including the straightforward extraction of relevant quantities, such as the (sub-)population net survival and associated intervals. This is true for any of the smooths included in the model as well. The simulation study, covering four sample sizes and several DGPs, demonstrated the reliability and flexibility of the proposed model. In particular, we have observed low levels of bias and variance of the point estimates, as well as a good ability to recover the excess hazard shape, whilst maintaining low computational times with the fitting procedure taking between 1 and 30 seconds for datasets of up to $n = 5000$ observations. 

The two case studies using real population-based cancer data highlight the usability of our methodology, which allows for the inclusion of complex effects to answer challenging research questions. We explored socio-demographic inequalities and geographic disparities in net survival for three of the most common cancer types diagnosed in England. In a wider context, such kind of results are increasingly being used to formulate cancer control strategies and to prioritize cancer control measures \citep{APPGC:2017,DofH:2019}. Other uses of our methodology include, but are not limited to, the study of long-term trends in net survival to evaluate the effectiveness of national cancer plans after they have been implemented, by assessing their impact on survival \citep{exarchakou:2018}. In addition, the possibility of modeling spatial effects through the MRF approach facilitates the study of geographic disparities in cancer survival for different sets of relevant health geographies. We note that age-standardization techniques, i.e. re-weighting the estimates of net survival using a standard cancer population age-distribution, have not been applied to the net survival estimates presented in the case studies. We emphasize that when the interest lies in comparing levels of survival between different populations or over time within the same population, such techniques can be applied to avoid that comparisons are masked by differences in the age profiles of cancer patients, since for most cancers the cancer-specific hazard is age dependent \citep{Corazziari:2004}. This is not to be confused with the use of the standardized age at diagnosis as a covariate in the model specification, which is indeed done in the case studies.

Our contribution to the relative survival setting brings new research opportunities, for instance: i) extending the proposed methodology to model multivariate survival data; a potential direction for such development consists of following the copula link-based model proposed in \cite{Marra2019}; ii) the additive decomposition of the total (or overall) hazard in the relative survival framework, relies on the assumption that the two competing risks (associated with the cancer and with all other causes of death) remain independent during the entire follow-up period. Copula functions could be used to relax this assumption; iii) developing formal model selection tools for additive excess hazard regression models \citep{maringe:2019,rossell:2019}; and iv) extending the applicability of our methodology by implementing additional quantities of interest for cancer research \citep{belot:2019} in the \texttt{GJRM} package.

\newpage

\section*{Acknowledgments}
AE was partly supported by the Windsor Fellowship DeepMind Computer Science Scholarships and by the UCL Departmental Teaching Assistantship Scholarship. MQ was funded through the Cancer Research UK Population Research Committee Funding Scheme: Cancer Research UK Population Research Committee - Programme Award C7923/A29018. GM and RR were supported by the EPSRC grant EP/T033061/1 during the revision of the work which followed the first submission. 

Finally, we would like to thank the two anonymous reviewers, the associate editor and the editor for their well thought out suggestions which helped us improve and clarify several aspects of the article.

\bibliographystyle{plainnat}
\bibliography{references}  

\end{document}